\documentclass[oldversion]{aa} 
\usepackage{graphicx}
\usepackage{longtable}
\usepackage{txfonts}
\usepackage{rotating}
\usepackage{lscape}

\begin{document}

\title{A vigorous activity cycle mimicking a planetary system in {\object HD~200466}
\thanks{Based on observations made with the Italian Telescopio Nazionale Galileo (TNG) operated 
on the island of La Palma by the Fundacion Galileo Galilei of the INAF (Istituto Nazionale di 
Astrofisica) at the Spanish Observatorio del Roque de los Muchachos of the Instituto de 
Astrofisica de Canarias}}

\author{ E. Carolo\inst{1,2},
         S. Desidera\inst{1},
         R. Gratton\inst{1},         
         A.F. Martinez Fiorenzano\inst{3},
         F. Marzari\inst{4},
         M. Endl\inst{5},
         D. Mesa\inst{1},
         M. Barbieri\inst{1},
         M. Cecconi\inst{3},   
         R.U. Claudi\inst{1},
         R. Cosentino\inst{3,6},      
         \and
         S. Scuderi\inst{6}}
\authorrunning{E. Carolo et al.}
\offprints{E. Carolo,  \\
   \email{elena.carolo@oapd.inaf.it} }
\institute{INAF -- Osservatorio Astronomico di Padova,  
             Vicolo dell'Osservatorio 5, I-35122, Padova, Italy
             \and 
             Dipartimento di Astronomia -- Universit\'a di Padova, Vicolo
             dell'Osservatorio 2, Padova, Italy 
             \and
             Fundaci\'on Galileo Galilei - INAF,
             Rambla Jos\'e Ana Fernandez P\'erez, 7
             38712 Bre\~na Baja, TF, Spain
             \and 
             Dipartimento di Fisica -- Universit\'a di Padova, Via Marzolo 8,
             Padova, Italy 
             \and
             McDonald Observatory, The University of Texas at Austin, Austin, 
             TX 78712, USA
             \and
             INAF -- Osservatorio Astrofisico di Catania, Via S.Sofia 78, Catania, Italy}

\date{Received  / Accepted }

\abstract{
Stellar activity can be a source of radial velocity (RV) noise and can reproduce periodic RV 
variations similar to those produced by an exoplanet. We present the vigorous activity cycle 
in the primary of the visual binary {\object HD~200466}, a system made of two almost identical 
solar-type stars with an apparent separation of 4.6~arcsec at a distance of $44\pm 2$~pc. High 
precision RV over more than a decade, adaptive optics (AO) images, and abundances have been 
obtained for both components. A linear trend in the RV is found for the secondary. We assumed 
that it is due to the binary orbit and once coupled with the astrometric data, it strongly 
constrains the orbital solution of the binary at high eccentricities ($e \sim$0.85) and quite 
small periastron of $\sim 21$~AU. If this orbital motion is subtracted from the primary radial 
velocity curve, a highly significant (false alarm probability $ < 0.1$\%) period of about 
1300~d is obtained, suggesting in a first analysis the presence of a giant planet, but it 
turned out to be due to the stellar activity cycle. Since our spectra do not include the Ca~II 
resonance lines, we measured a chromospheric activity indicator based on the $H_\alpha$ line to 
study the correlation between activity cycles and long-term activity variations. While the 
bisector analysis of the line profile does not show a clear indication of activity, the correlation 
between the $H_\alpha$ line indicator and the RV measurements identify the presence of a strong 
activity cycle.
}
          

   \maketitle
%

\section{Introduction}
\label{s:intro}

The impact of star spots on high precision RV surveys on rotational timescales is well 
documented by several observational detections collected in the past years (see e.g.
Queloz et al. 2001 and Paulson et al. 2004). Various authors 
explained such observations by means of theoretical modelling (Saar \& Donahue 1997; 
Hatzes 2002; Desort et al. 2007). On the other hand, the impact of 
solar-like activity cycles on high-precision RVs has been studied only recently by analysing 
the correlations between observed jitter and chromospheric activity for the stars in the 
California Planet Search (Isaacson \& Fischer 2010) or by computing 
chromospheric activity indicators based on some element lines as in the HARPS program (Gomes da 
Silva et al.~2011). Some authors proposed several methods and models 
to reduce the associated noise (Dumusque et al. 2011, Meunier \& Lagrange 2013) and predict the 
activity-induced RV variations (Lanza et al. 2011).

Active regions are known to affect the convection pattern in the Sun, with effects on line 
bisectors and line shifts, possibly mimicking the low-amplitude RV variations due to 
extrasolar planets (Dravins 1992).

Meunier et al.~(2010) and Meunier \& Lagrange (2013) used the spots and plages
observed on the Sun over a 
solar cycle to derive the expected RV signature. They considered both the photometric 
contribution of spots and plages and the contribution of the latter due to the attenuation 
of the convective blueshift, which is the dominant source of RV variability, with 
peak-to-valley amplitude of about 8-10 m/s over a solar cycle. Boisse et al. (2011) 
simulated dark spots on a rotating stellar photosphere in order to better understand and 
characterize the effects of stellar activity and to differentiate them from planetary 
signals. Their approach was also applied to and validated on four active planet-host stars. 
Useful results were obtained, providing some constraints on the planet period, on semi-amplitude 
of the planet, on star rotational period, and on data covering. Saar \& Fischer (2000) 
looked for correlations between the strength of the Ca~II infrared triplet and RV in
the Lick planet search data and found significant correlations in about 30\% of the stars.
They also aimed at correcting the RVs for the effects of magnetic activity using these
correlations. They found that less active stars with significant Ca~II variations are those
that are best corrected. These variations were explained as due to long-term
activity cycles, while corrections become less reliable at increasing activity levels when 
short-term noise due to star-spots becomes the dominant source of RV variations.
It should be noticed that plages affecting convection do not only produce a RV signal 
on long-term scales (cycle), but also produces a modulation due to rotation period, on 
smaller timescales.

In Isaacson \& Fischer (2010) the $\Delta S$ index of excess emission in the
core of Ca II $H\&K$ lines was defined and correlated to the radial velocity jitter, as a 
function of the star $B-V$ colour. They found that chromospherically quiet F and G dwarfs 
and subgiants exhibit higher baseline levels of the astrophysical jitter than K dwarfs. 
They also concluded that the correlation between the activity cycle and the Doppler 
measurements is rare to see, making the correction of activity-induced velocity 
variations in F and G dwarf difficult.

Recently, Lovis et al.~(2011) studied the magnetic activity cycles 
in solar-type stars in the HARPS program analysing the Ca II $H\&K$ lines. They found 
that $39\%$ of the old solar-type stars in the solar neighbourhood do not show any 
activity cycles, while $61\%$ do have one. They also confirmed that the sensitivity of 
radial velocities to magnetic cycles increases towards hotter stars, while late K dwarfs 
are almost insensitive (confirming the result of Isaacson \& Fischer 2010). 
For the HARPS sample they concluded that the activity cycles can induce radial velocity 
variations having long period and amplitude up to about 25 m/s.

From the line analysis on a sample of M dwarf stars from the HARPS program, about $40\%$ 
of the stellar sample showed significant variability. Gomes da Silva et 
al.~(2011) concluded that $H_\alpha$ shows good correlation with 
$S_{\rm Ca~II}$\ for the most active stars, indicating that the correlation between Ca II and 
$H_\alpha$ depends on the activity level of the star. The correlation between different
activity indicators is complex and can be different depending on the star (Cincunegui 
et al. 2007; see also Meunier \& Delfosse 2009).

Therefore care should be taken in the claim of detection of planets around active stars. 
False alarms have been reported in the past
(see e.g. Hernan-Obispo et al. 2010; Figueira et al.~2010; Setiawan et al. 2008; Huelamo et al.~2008) 
and controversial cases for which conclusive evidence of the Keplerian origin of the RV variations
is still lacking have been discussed in e.g. Hatzes \& Cochran (1998).
The measurements of activity indicators and line profile variations on the same spectra used
for the determination of high-precision RVs are crucial to disentangling RV variations due
to Keplerian motions from those related to magnetic activity.
Calcium II H\&K lines represent the most widely used chromospheric activity indicator for solar-type stars. 
For these stars, $H_\alpha$ has been mostly neglected but it is a suitable alternative
when Ca II H\&K  lines are not included in the spectra used for RV determination.
Line profiles are also sensitive to phenomena related to magnetic activity. 
Furthermore, they are a unique diagnostic for checking
the occurrence of contamination in binary systems (Martinez Fiorenzano et al.~2005; Torres et al.~2004).

Within our survey looking for giant planets in nearly equal mass visual binaries with the 
SARG spectrograph at Telescopio Nazionale Galileo (TNG) (Desidera et al. 2010), we discovered 
an interesting system (HD~200466) where the primary exhibits large variations of the radial 
velocities that are correlated with variations of activity indicators, while the secondary 
only shows a long-term trend in radial velocities. While almost twin late-G dwarfs, 
the two stars also 
have other differences, markedly in the photospheric Li abundance. This paper presents the 
observations of this system. The paper is organized as follows. In Sect.~\ref{s:obs} we 
describe the observations and reduction of the spectroscopic and adaptive optics data. We 
also present the analysis of the $H_\alpha$ line performed to detect the presence of stellar 
activity. In Sect.~\ref{s:param} we describe the stellar parameters of each component. In 
Sect.~\ref{s:rv} we present the radial velocity variations for both components. In 
Sect.~\ref{s:bin} we present the clues of the binary orbital solution and the plausible 
system architecture. In Sect.~\ref{s:rvfit} we assume and analyse the presence of a planet 
around the primary component. In Sect.~\ref{s:origrv} we argue that the RV variations 
observed in {\object HD~200466}A are related to the stellar activity cycle. 
In Sect.~\ref{s:correction} we consider approaches to correct the RV time series for the
effects of chromospheric activity.
In Sect.~\ref{s:discussion} we discuss the results in the context of the possible effect
of the presence of planets on the lithium content of solar-type stars and summarize our 
conclusions.

\section{Observations and data reduction}
\label{s:obs}

\subsection{High resolution spectroscopy with SARG}

Spectroscopic observations were performed using SARG, the high resolution spectrograph 
of TNG (Gratton et al.~2001). The instrument set up is described in 
Desidera et al.~(2011). Integration time was fixed at 900 s to keep errors 
due to the lack of knowledge of flux mid-time of each exposure below photon noise. Typical S/N ratio at 
about 5800~\AA~is about 100 per pixel. Overall, 76 and 72 spectra of {\object HD~200466} 
A and B, respectively, were acquired from September 2000 to September 2011.

\begin{figure}
\includegraphics[width=9cm]{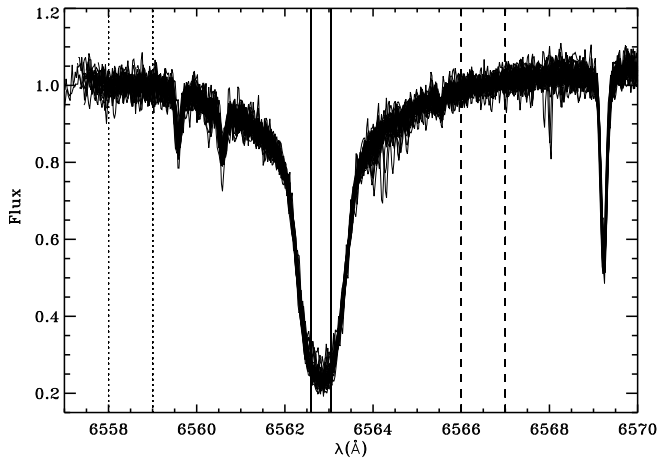}
\caption{$H_{\alpha}$ profile of a sample of 30 spectra of {\object HD~200466}A. The dotted lines 
correspond to the two ranges used to calculate the normalized continuum. The solid line 
corresponds to the ``on line'' band used for the activity analysis. }    
\label{f:halphalineA}
\end{figure}

Data reduction was performed in a standard way using IRAF. However for the $H_\alpha$ 
analysis we used an ``ad hoc'' continuum normalization of spectra, because this broad 
line is located close to the blue edge of the order 93 of SARG echelle spectra. Since 
the spectral line is very broad, the standard IRAF procedure to fit a polynomial 
through the highest spectral point is not adequate. In order to extract information 
on this line flux we then calculated the continuum separately from the standard IRAF 
data reduction (see Fig.~\ref{f:halphalineA}). 

Radial velocities were obtained using the AUSTRAL code (Endl et al.~2000). 
Internal errors of radial velocities are typically about 4 m/s. Tables \ref{t:rva} and 
\ref{t:rvb} list the radial velocities for {\object HD~200466} A and B, respectively. 

Average line profiles were also derived to get additional diagnostics of spurious RV 
variations due to stellar activity, contamination, or other causes. We followed the 
approach developed in Martinez Fiorenzano et al.~(2005), but considering a more 
extended line list with respect to previous analysis to build the mask involved in the 
CCF calculation. The top and bottom zones for the calculations of the bisector velocity 
span (BVS) were set as follows: top centred at 25\% of the maximum absorption, bottom 
at 87\%; in both cases the width was of 25\%. The BVS are listed in Cols. 4 and 5 of 
Tables \ref{t:rva} and \ref{t:rvb}, with their errors.

The analysis of the $H_\alpha$ line was performed to detect the presence of stellar 
activity. The approach used is based on measuring an ad hoc flux index in the line 
as well as in adjacent continuum bands on both the short and long wavelength wings for 
each spectra of both companions (see e.g. Fig.~\ref{f:halphalineA}). The values for this
index are also listed in Cols. 6 and 7 of Tables \ref{t:rva} and \ref{t:rvb}, with 
their errors.

\subsection{Direct imaging with AdOpt@TNG}

\subsubsection{Observations}

To complement the spectroscopic data, {\object HD~200466} was observed on July 12 and 
20, 2007, and Aug 9, 2008, with AdOpt@TNG, the adaptive optics module of TNG (Cecconi et 
al.~2006). The instrument feeds 
the HgCdTe Hawaii 1024x1024 detector of NICS, the near infrared camera, and spectrograph 
of TNG, providing a field of view (FoV) of about 44 $\times$ 44 arcsec, with a pixel scale of 
0.0437 \arcsec/pixel. Plate scale and absolute detector orientation were derived
as part of the program of follow-up of systems shown to have long-term radial velocity 
trends from the SARG planet search (see Desidera et al.~2010 for some
preliminary results). We did not find significant variations of the plate scale with time, 
while the angular offset between the nominal detector orientation and the true direction 
of north was found to have small but significant differences between the 2007 and 2008 
seasons, likely due to the refurbishment of the instrument made during the 2007-2008 
winter. As our program with AdOpt@TNG is focused on the follow-up of stars that show
long-term radial velocity trends, in several cases we expect the occurrence of 
detectable astrometric variations induced by the companion. The final value of the plate 
scale depends very little on the choice of the subsample used to derive it (all targets 
or only the few systems without large RV trends), but the dispersion is instead much 
larger when all the targets are included. Furthermore, we are working on the 
determination and correction of the optical distortions. At the three observing epochs 
we acquired 118, 134, and 98 images on {\object HD~200466} in Br$\gamma$ 
intermediate-band filtre, avoiding detector saturation. At each epoch, observations with 
different instrument rotation images were obtained. This makes the optimization 
of the position shift between the images of the components easier and provides meaningful
results as close as possible to the stars. The primary was used as the reference star for 
the adaptive optics. The target was observed in different positions on the detector for 
sky subtraction purposes, and at two or three orientations of the FoV on each run to
allow a better disentangling between true companions and instrumental artefacts.

\subsubsection{Data analysis procedure}

\begin{figure}
\includegraphics[width=9cm]{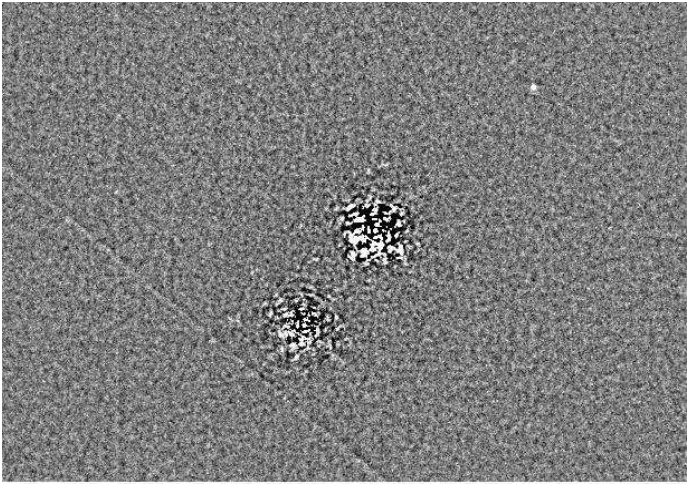}
\caption{Residuals on the composite AdOpt@TNG image after subtraction of the two 
components of the {\object HD~200466} system}    
\label{f:image}
\end{figure}

\begin{figure}
\includegraphics[width=9cm]{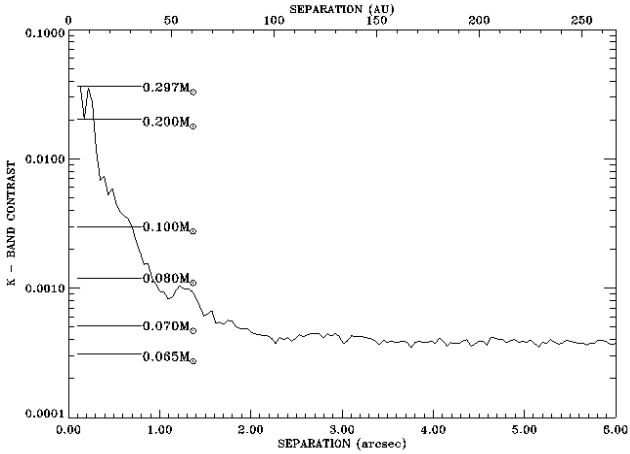}
\caption{Detection limits from AdOpt@TNG images of {\object HD~200466}. The contrast 
in the $Br \gamma$ band is shown as a function of projected separation in arcsec and AU. 
Corresponding values of mass limits are shown.}    
\label{f:limitsadopt}
\end{figure}

The data analysis procedure is described in detail in Desidera et al. (2011) 
and we briefly summarize it here. We first corrected for detector cross-talk using the 
dedicated code developed at TNG\footnote{\tt www.tng.iac.es/instruments/nics/files/crt\_nics7.f\,.}.
Then we performed standard image preprocessing (flat fielding, bad pixels, and sky 
corrections) in the IRAF environment. In the data analysis, we exploit the similarity 
between the point spread functions (PSFs) of the two components and 
the availability of observations taken at different rotation angles. Datacubes of images 
taken with the same rotation angles were built and a median combined image was obtained, 
discarding images with large FWHM. The resulting images for the three orientations were 
then rotated and summed together. To enhance detectability of faint companions at small 
projected separation, we considered the square regions in the final image 
composed of $51\times 51$\ pixels around the position of the two components, and
subtracted each of them to the other after scaling
them for the flux difference. This PSF subtraction procedure works quite well because the 
two stellar images are well within the isoplanatic angle. It allows a gain of about an 
order of magnitude in contrast at separation $<1$~arcsec. Figure~\ref{f:image} shows the 
image after the subtraction of the two components of the {\object HD~200466} system. We 
did not find any evidence of faint companions to either of the two components. In 
Figure~\ref{f:limitsadopt} we display the limiting contrast obtained around the B component and 
the corresponding mass detection limits (see next paragraph). The limiting contrast was obtained considering 5 times the standard 
deviation in circular annuli around the calculated position of the star. During these 
calculations the A component was masked. We verified that the contrast curve agreed closely
with a noise model which included speckle, photon, sky background, and detector noise.
These detection limits were further validated by injecting a number of fake objects at 
different separations and at different contrast and performing the same analysis as on 
the original images.

To obtain the mass detection limits from the contrast limits we first transformed the 
luminosity contrast  in absolute magnitude in the K band. For $M_{\rm K}$ magnitudes brighter 
than $9.5$\ we used the mass-luminosity relation given in Delfosse et 
al.~(2000) for low mass stars ($M<0.6M_{\odot}$), obtaining a mass value for 
each magnitude limit. For $M_{\rm K}$\ in the range of $9.5<M_{\rm K}<12.8$\ we interpolated the 
tables by Chabrier et al.~(2000) for the star age derived in 
Section~\ref{s:param} (2 Gyr) while for magnitudes fainter than $M_{\rm K}~12.8$ we used the 
same method but using the tables from Baraffe et al.~(2003)\footnote{The transition 
between the two sets of models corresponds to an effective temperature of about 1500~K, 
which roughly separates the validity ranges of dusty models (Chabrier et al.~2000) 
and COND models (Baraffe et al.~2003).}.
To smooth the irregularities of three range of masses we fitted the distribution by using 
an exponential function reproducing the mass detection limits for each separation.

\section{Stellar parameters}
\label{s:param}

For a proper interpretation of the observed radial velocity variability, a careful 
evaluation of the stellar properties is mandatory. Table \ref{t:starparam} summarizes 
the stellar parameters of the components of {\object HD~200466}, that are described 
in more detail below.

\begin{table}[h]
\caption{Stellar properties of the components of {\object HD~200466}}
\smallskip
\begin{minipage}{8.8cm}
\begin{tabular}{lccc}
\hline
Parameter   &  HD~200466A &  HD~200466B  & Ref. \\
\hline
$\alpha$~(2000)         & 21 02 21.905    & +37 39 13.93           & 1 \\
$\delta$~(2000)         & 21 02 21.514    & +37 39 13.54           & 1 \\
$\mu_{\alpha}$~(mas/yr) & \multicolumn{2}{c}{-78.54 $\pm$ 0.82}    & 2 \\
$\mu_{\delta}$~(mas/yr) & \multicolumn{2}{c}{-221.29 $\pm$ 1.07}   & 2 \\
RV~(km/s)(2001)         & $-8.19\pm0.20$  & $-8.44\pm0.20$         & 3 \\
$\Delta$ RV (B-A)~(m/s) (2007)& \multicolumn{2}{c}{-436 $\pm$50 }        & 3 \\
$\pi$~(mas)             & \multicolumn{2}{c}{22.53 $\pm$ 1.01}     & 2 \\
$\pi$~(mas)             & \multicolumn{2}{c}{22.83 $\pm$ 1.75}     & 1 \\
$d$~(pc)                & \multicolumn{2}{c}{ $44.4^{+2.1}_{-1.9}$}& 2 \\
$U$~(km/s)              & \multicolumn{2}{c}{ 43.73$\pm$1.44 }     & 3 \\
$V$~(km/s)              & \multicolumn{2}{c}{-17.34$\pm$0.35 }     & 3 \\
$W$~(km/s)              & \multicolumn{2}{c}{-17.27$\pm$0.97 }     & 3 \\
$R_{min}$~(kpc)         & \multicolumn{2}{c}{6.856 $\pm$  0.031 }  & 3 \\
$R_{max}$~(kpc)         & \multicolumn{2}{c}{9.773 $\pm$  0.051 }  & 3 \\
$z_{max}$~(kpc)         & \multicolumn{2}{c}{0.125 $\pm$  0.013 }  & 3 \\
$e_{m}$                 & \multicolumn{2}{c}{0.175 $\pm$  0.004 }  & 3 \\
\hline
Spectral Type    	    &       K05         &      K05             & 6 \\
V                       & 8.399 $\pm$ 0.005 & 8.528 $\pm$ 0.006    & 4 \\
B-V                     & \multicolumn{2}{c}{0.748$\pm$0.012}      & 1 \\
B-V                     & 0.71  $\pm$ 0.02  & 0.79 $\pm$ 0.03      & 4 \\
B-V~(calc)              & 0.736             & 0.761                & 3 \\
V-I                     & \multicolumn{2}{c}{0.79$\pm$0.01}        & 1 \\
$H_{p}$                 &  8.546$\pm$ 0.005 &  8.687 $\pm$ 0.006   & 1 \\
$H_{p}$~scatter         &\multicolumn{2}{c}{0.012\footnote{A+B}\footnote{See Section \ref{s:variability} for details}}& 1 \\
J                       & 6.916 $\pm$ 0.041 & 6.942 $\pm$ 0.027    & 5 \\ 
H                       & 6.603 $\pm$ 0.038 & 6.586 $\pm$ 0.049    & 5 \\
K                       & 6.574 $\pm$ 0.036 & 6.556 $\pm$ 0.029    & 5 \\
\hline
$M_{V}$                 & 5.16$\pm$0.10 &   5.29$\pm$0.10          & 3 \\
$T_{eff}$ (K)           & 5604 $\pm$45   &  5551 $\pm$45           & 3 \\
$\Delta T_{eff}(A-B)$ (K)& \multicolumn{2}{c}{53 $\pm$ 23}         & 3 \\
$\log g$                & 4.45 $\pm$ 0.10 & 4.47 $\pm$ 0.10        & 3 \\
${\rm [Fe/H]}$          & +0.05 $\pm$ 0.10 & +0.03 $\pm$ 0.10      & 4 \\
$\Delta {\rm [Fe/H]}(A-B)$& \multicolumn{2}{c}{0.020 $\pm$ 0.024 } & 4 \\
$\log N_{Li}$           & 1.2 & 2.17                               & 3  \\
\hline
S Index (1998)          &  0.221    &  0.258                       & 3 \\
$\log R^{'}_{HK}$ (1998)& -4.77     &  -4.69                       & 3 \\
$ v \sin i $~(km/s)     & 1.2 $\pm$0.5   & 2.2 $\pm$0.8            & 3 \\
$L_{X}$~(erg/s)         & \multicolumn{2}{c}{$1.4~10^{28~\mathrm{a}}$} & 3 \\
\hline
${\rm Mass} (M_{\odot})$& 0.947$\pm$0.035 & 0.929$\pm$0.033        & 3 \\
${\rm Radius} (R_{\odot})$& 1.0 & 1.0                              & 3 \\ 
Age~(Gyr)               &\multicolumn{2}{c}{$\sim 2.0$}            & 3 \\
\hline
\end{tabular}
References: 1 Hipparcos (ESA 1997); 2 Van Leuween (2007); 3 Carolo (2012);
4 Desidera et al. (2004); 5 Skrutskie et al. (2006); 6 Nassau \& Stephenson (1961)
\par
\vspace{-0.75\skip\footins}
\renewcommand{\footnoterule}{}
\end{minipage}
\label{t:starparam}
\end{table}

\subsection{Spectroscopic and photometric parameters}
\label{s:photom}

The abundance analysis performed in Desidera et al.~(2004) relies on the absolute 
magnitudes for the derivation of stellar gravity (and then indirectly on the determination 
of effective temperatures using the ionization equilibrium). In that paper, we used the HIPPARCOS 
parallax (ESA 1997). We adopted the revised parallax distance published by Van 
Leuween (2007) and used the web interface of Padova stellar models 
{\em param} \footnote{\tt http://stev.oapd.inaf.it/param } (Da Silva et al. 2006) 
to derive stellar masses in a fully self-consistent way. Then we repeated the abundance 
analysis. This produced results similar to the previous one (slightly cooler effective 
temperatures, and very similar temperature and abundance differences). The final 
atmospheric parameters and the adopted stellar masses are listed in Table \ref{t:starparam}.

As this analysis uses the trigonometric parallax and the stellar models, it cannot be 
considered an independent check of the reliability of the parallax and of the absolute 
effective temperature. To this purpose, we consider photometric colours and alternative
spectroscopic determinations of effective temperatures.

Few high quality measurements of magnitudes of the two components are available in the 
literature. Combining them, Desidera et al.~(2004) derived $V_A=8.399 \pm 0.005$, 
$V_B=8.528 \pm 0.006$, $B-V_A=0.71 \pm 0.02$, and $B-V_B=0.79 \pm 0.03$ mag. The colour 
difference of Tycho photometry for the two components is much larger than that expected from the magnitude
($\Delta V=0.13$) and spectroscopic temperature differences ($\Delta T_{\rm eff}=53$~K). 
Similar discrepancies are common for the pairs of our sample (see Desidera et al.~2004, 
their Table 5). It is then likely that internal errors in Tycho colours for individual
components have been underestimated for close binaries.

The {\em V-K} colours resulting from 2MASS photometry (Skrutskie et al.~2006) are
probably not very accurate. They put both the 
components to the red of the main sequence for the appropriate metallicity by about 
0.2~mag; furthermore, the secondary is marginally brighter in {\em K}. This is likely 
due to the uncertainties in 2MASS photometry which are much larger than the nominal ones 
for bright close binaries (aperture photometry on an unresolved object). Our AdOpt images 
yield a magnitude difference in the $Br \gamma$ filtre of 0.10 mag (with the primary being 
brighter), as expected from the {\em V} mag difference\footnote{This is further evidence
against the presence of additional bright stellar components in the system.}. As the resolved 
colours have significant errors, we consider composite photometry, that is available with 
high accuracy in the Hipparcos catalogue ($(B-V)_{A+B}=0.748\pm0.012$, $(V-I)_{A+B}=0.79\pm0.01$).
From the composite colour and the observed V mag difference, assuming the stars follow a 
standard isochrone, we derived $(B-V)_{A}=0.736$, $(B-V)_{B}=0.761$, $(V-I)_{A}=0.778$, and 
$(V-I)_{B}=0.801$. These V-I and B-V colours are only slightly redder than the isochrone and 
would require shifts of just 0.05 and 0.10 mag in $M_{V}$ for $(B-V)$ and $(V-I)$ colours,
respectively.

A fully spectroscopic analysis, deriving $T_{\rm eff}$ from excitation equilibrium, yields
temperatures that are cooler by about 120 K than those from ionization equilibrium (assuming 
nominal parallax) (5486 vs 5603~K for {\object HD~200466}A), corresponding to a 0.20~mag 
shift in $M_{V}$. All of this suggests that the trigonometric parallax is basically correct, 
although the system might be slightly closer than 44.4 pc. A distance closer than 40 pc is
 unlikely.

\subsection{Lithium}
\label{s:li}

\begin{figure}
\includegraphics[width=9cm]{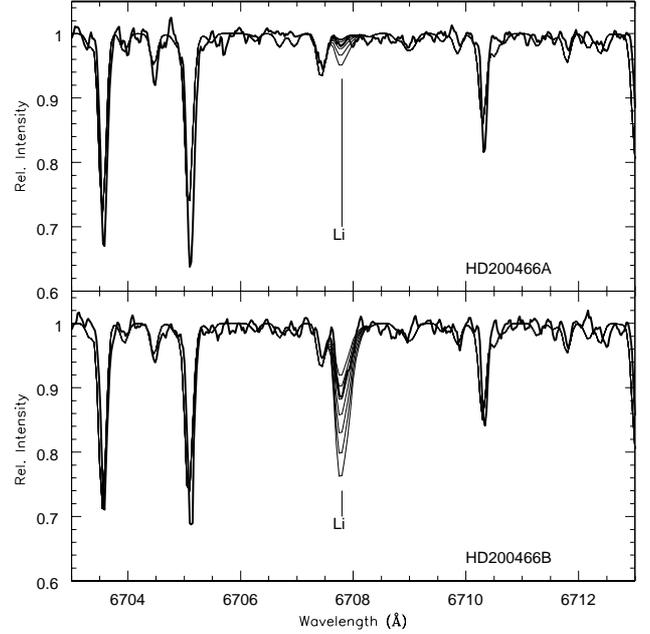}
\caption{Portion of spectra of HD~200466A (upper panel) and B (lower panel) close to Li 
6708\AA~(thick solid line)
overplotted with the synthetic spectra (thin solid lines). They are derived for:
log n(Li)=0.6, 0.8, 1.0, 1.2, 1.4, 1.6, 1.8 for {\object HD~200466}A and
log n(Li)=2.0, 2.1, 2.2, 2.3, 2.4, 2.5, 2.6 for HD~200466B}        
\label{f:li}
\end{figure}

Visual analysis of the spectra of {\object HD~200466} A and B shows a significant 
difference in the 6708~\AA~Li doublet. We performed spectral synthesis of this 
spectral region to determine the Lithium abundance. We adopted the atmospheric 
parameters resulting from the updated abundance analysis (Fig.~\ref{f:li}).
We found $\log N_{Li}=1.2 \pm 0.1 $ for {\object HD~200466}A and $\log N_{Li}=2.17 
\pm 0.1 $ for {\object HD~200466}B. Detection of Li in the primary is 
marginal; values lower than this are therefore possible.

\begin{figure}
\includegraphics[width=9cm]{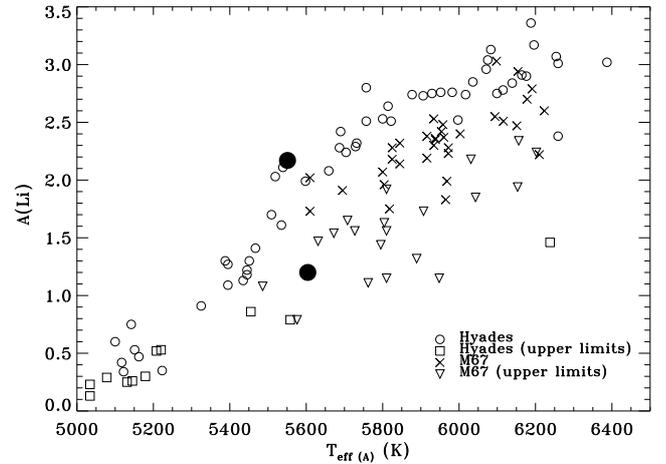}
\caption{Li abundance vs effective temperature for the components of HD~200466A (lower black point)
and B (upper black point) and
stars of Hyades (640 Myr) and M67 (4 Gyr).}        
\label{f:liabu}
\end{figure}

The Li content of component A is well below that of Hyades stars of similar $T_{\rm eff}$
and comparable to that of Li-poor stars in M67, while that of {\object HD~200466}B 
is comparable to that of Hyades (Fig.~\ref{f:liabu}).
The smaller Lithium abundance of {\object HD~200466}A with respect to 
{\object HD~200466}B cannot be due to the temperature difference, as larger depletions 
are expected at colder temperatures. Instead, this is the signature of significant 
intrinsic differences in the lithium depletion history between the two components.

\subsection{Stellar age}
\label{s:age}

Given the main sequence status of both components and the temperatures colder than the Sun, 
isochrone fitting (see above) gives inconclusive results for the ages ($3.6 \pm 3.4$ Gyr 
for the primary and $3.7 \pm 3.4$ Gyr for the secondary). As shown in Sect.~\ref{s:li}, 
the lithium abundance of B is similar to Hyades stars of similar temperature, while that of A 
is well below and similar to older stars. We then considered additional age indicators, 
based on magnetic activity, rotation, and kinematics.

\subsubsection{X-ray emission}

ROSAT identified a source (\object{1RXS J210223.3+373856}) close to {\object HD~200466} 
(separation 24 arcsec with a quoted error of 33 arcsec) (Voges et al.~2000).
Assuming this is the X-ray counterpart of HD200466, the
X-ray luminosity derived using the calibration of H\"unsch et al.~(1999) 
is $L_{X}=1.40~10^{28}$~erg/s (A+B components). Assuming similar luminosities of both 
components, we obtain $L_{X}=7.0~10^{27}$~erg/s for the individual stars. The age 
resulting from the calibration by Mamajek \& Hillebrand (2008) is 2.9 Gyr.
 
\subsubsection{Chromospheric emission}

The adopted SARG set-up does not include Ca II H\&K in the spectral format. However, 
the components of {\object HD~200466} were observed three times in 1998 with HIRES at Keck
as part of the G Dwarf Planet Search (Latham 2000). We retrieved the reduced 
spectra from Keck archive\footnote{If we considered the phase and length of the
activity cycle discussed in Section 7, these spectra were taken close to a maximum in
the activity cycle. However, activity data also show a long-term trend
that if extrapolated at the epoch of these observations indicates an activity
level well below the average of our observation.}.

To quantify the chromospheric emission, we built a calibration of the S index as 
measured in the Keck spectra and in the standard M.~Wilson system. To this end we 
retrieved from the Keck archive several spectra of 15 stars from the list of Wright et 
al.~(2004) in the colour range $0.72< B-V <0.76$ and spanning various activity levels.
The calibration into the standard Mt.~Wilson system has a dispersion of 0.011, likely 
dominated by intrinsic variability of the chromospheric activity.

The values of $\log R_{HK}$ of {\object HD~200466}A and B derived following the 
prescriptions of Noyes et al.~(1984) and for the calculated individual 
B-V colours were found to be of  $-4.77$ and $-4.69$ for {\object HD~200466}A and B,
respectively.
The ages resulting from the calibration by Mamajek \& Hillebrand (2008)
are 2.8 and 1.9 Gyr for {\object HD~200466}A and B, respectively. This difference
provides an estimate of the rather large uncertainties related to this method.


\subsubsection{Rotational period}
\label{s:variability}

%
%

\begin{figure}
\includegraphics[width=9cm]{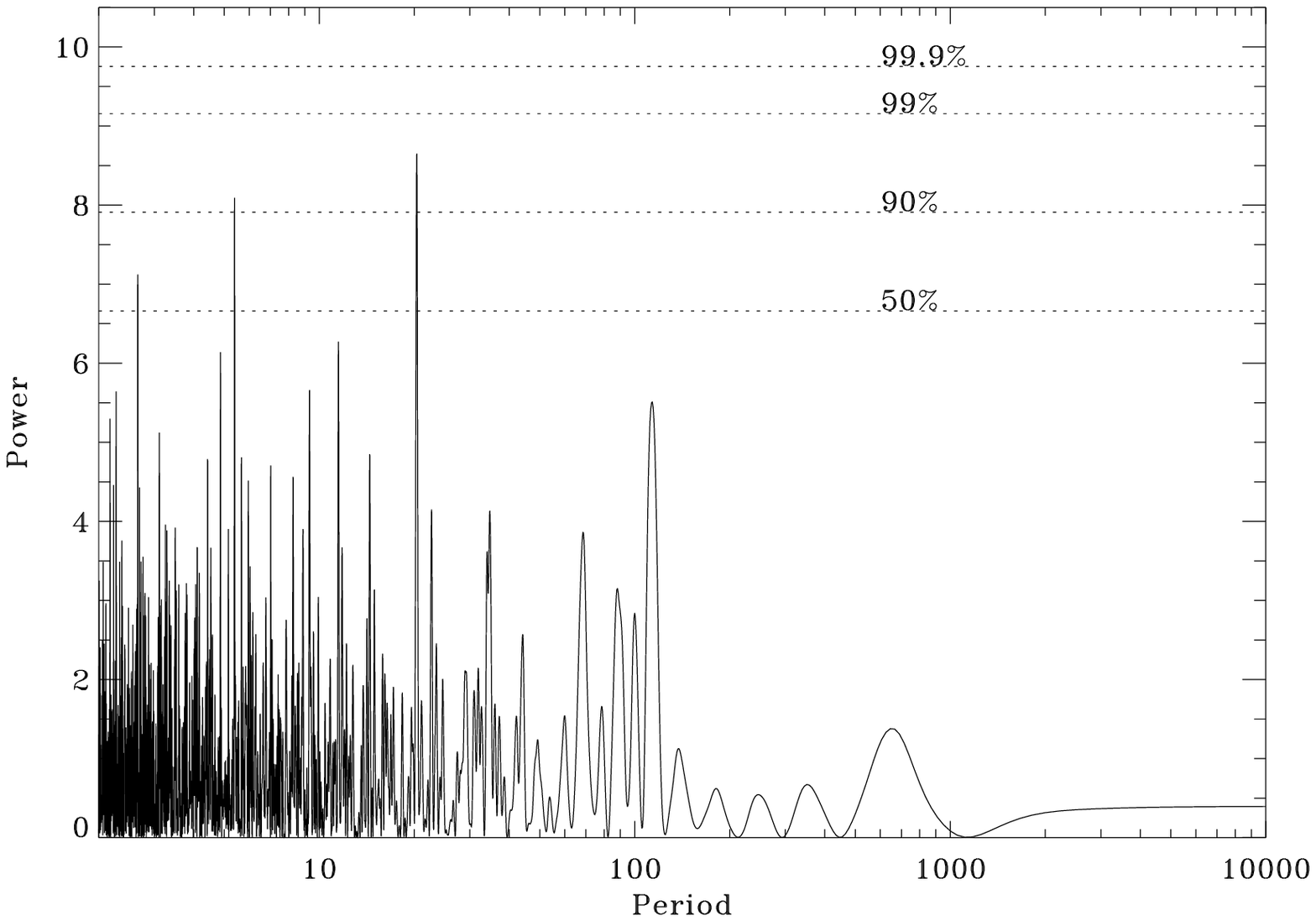}
\includegraphics[width=9cm]{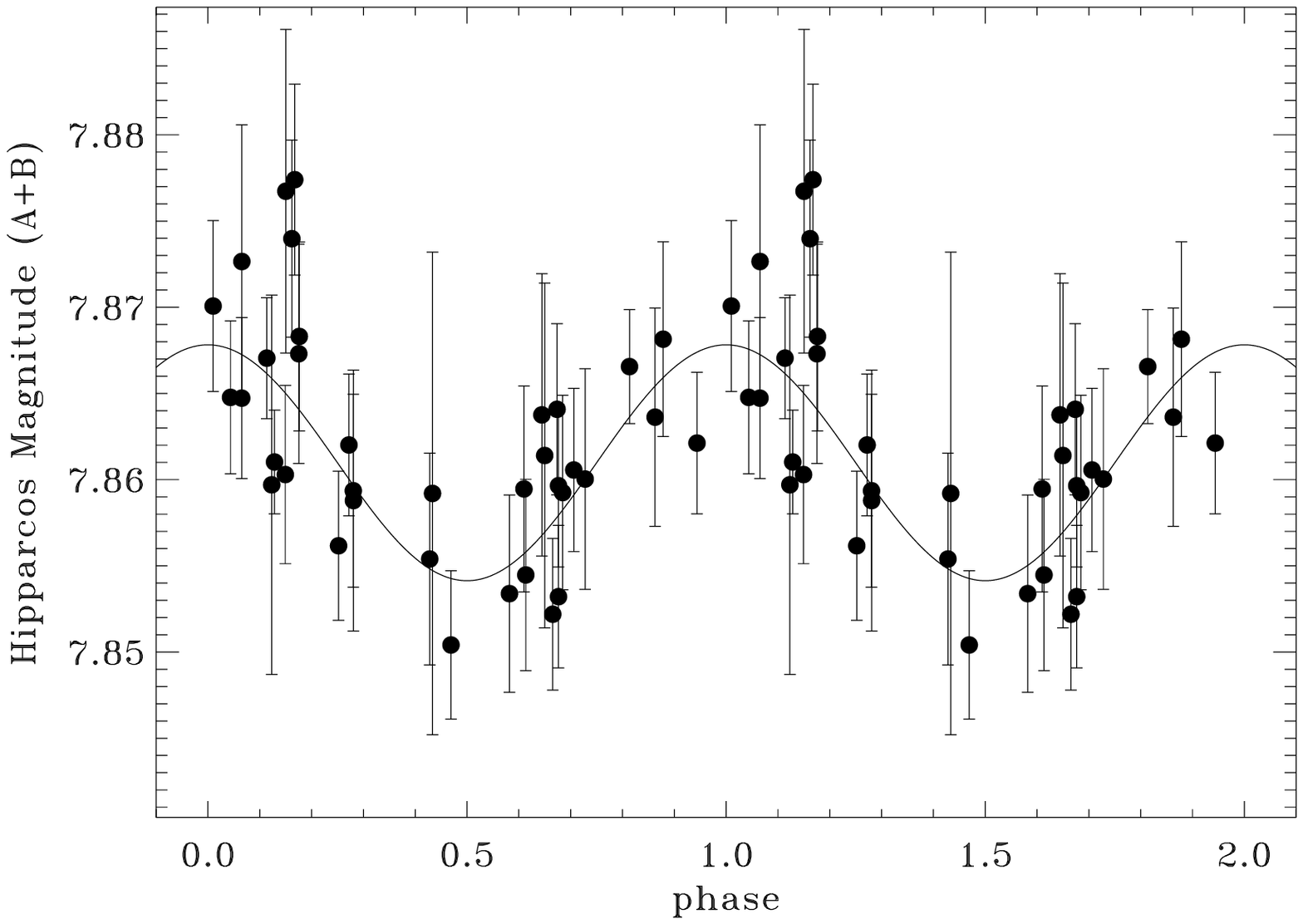}
\caption{Hipparcos photometry (joined A+B components). Upper panel: Lomb-Scargle periodogram;
Lower panel: photometry phased to the best-fit period of 20.34 d.}        
\label{f:hipphot}
\end{figure}

The only photometric time series available is that of Hipparcos. Hipparcos photometry 
refer to joined A+B components. The errors in the resolved photometry by Tycho are large.
We analysed the individual Hipparcos photometric data, eliminating one obvious outlier 
and performing daily averages. The dispersion of the daily averages is 0.007 mag.
The periodogram shows a peak at 20.34 days (Fig.~\ref{f:hipphot}), which has a false alarm 
probability of 2.5\% according to a bootstrap test. While not highly significant, it might 
represent the rotational period of one of the components. Indeed, the expected rotational 
period derived from the observed Ca II H\&K emission using the relation by Mamajek \& 
Hillebrand (2008) is just slightly longer (26.0 and 23.4 days for {\object HD~200466} A 
and B, respectively) and the observed period is compatible with the projected rotational 
velocities (see below) for inclinations of about 30 and 55 deg for HD~200466A and B, respectively.

While neither the photometric nor the spectroscopic periods have confidence levels
larger than 99\%, the detection of a very similar period with independent techniques and 
at different epochs is a strong indication that indeed the rotation period of HD~200466A is 
about 20 d.
The corresponding ages resulting from the calibration by Mamajek \& Hillebrand (2008) are 
2.3 and 2.1~Gyr depending on whether the 20.3~d period belongs to {\object HD~200466}A or B.

Projected rotational velocities were derived performing an FFT analysis of the 
CCF profiles (for all spectra involved in the line profile analysis). The procedure is 
described in Desidera et al.~(2011). The macroturbulence was estimated from effective 
temperature and the {\em B-V} colours following the calibration by Valenti \& Fischer
(2005). We found $v \sin i_{A}=1.3\pm0.5 $ km/s (upper limit) and $v \sin i_{B}=2.2\pm0.8 $ km/s.

\subsubsection{Kinematics}

Absolute radial velocities of the SARG templates of {\object HD~200466} A and B as derived using 
cross-correlation with spectra of a few objects from the list of Nidever et al.~(2002)
lead to $-8.19\pm0.20$ and $-8.44\pm0.20$~km/s, very similar to the values listed
in Carney et al. (1994: $-8.0$\ and $-8.4$~km/s) and Nordstrom et al. (2004: 
$-8.3$~km/s from composite spectra).

We use our absolute radial velocities (corrected for gravitational and convective shifts
as in Nidever et al.~2002) together with Hipparcos astrometry to derive
space velocities and galactic orbit as in Barbieri \& Gratton (2002).
The space velocity is far from the regions typical of young stars (see e.g. Montes
et al.~2001).
Overall, the kinematics is fully compatible with an age of 2-4 Gyr.

\subsubsection{Summary of stellar ages}

In summary, {\object HD~200466} is composed of two very similar components with masses
slighly lower than and iron content similar to or slightly enriched with respect to the 
Sun. The isochrone fitting is inconclusive with regard to stellar age. From lithium, 
Ca II H\&K emission, X ray coronal emission, and the tentative rotational period we 
derived a most probable age of about 2 Gyr.

\section{Radial velocity variations of HD~200466A and B}
\label{s:rv}

\begin{figure}
\includegraphics[width=9cm]{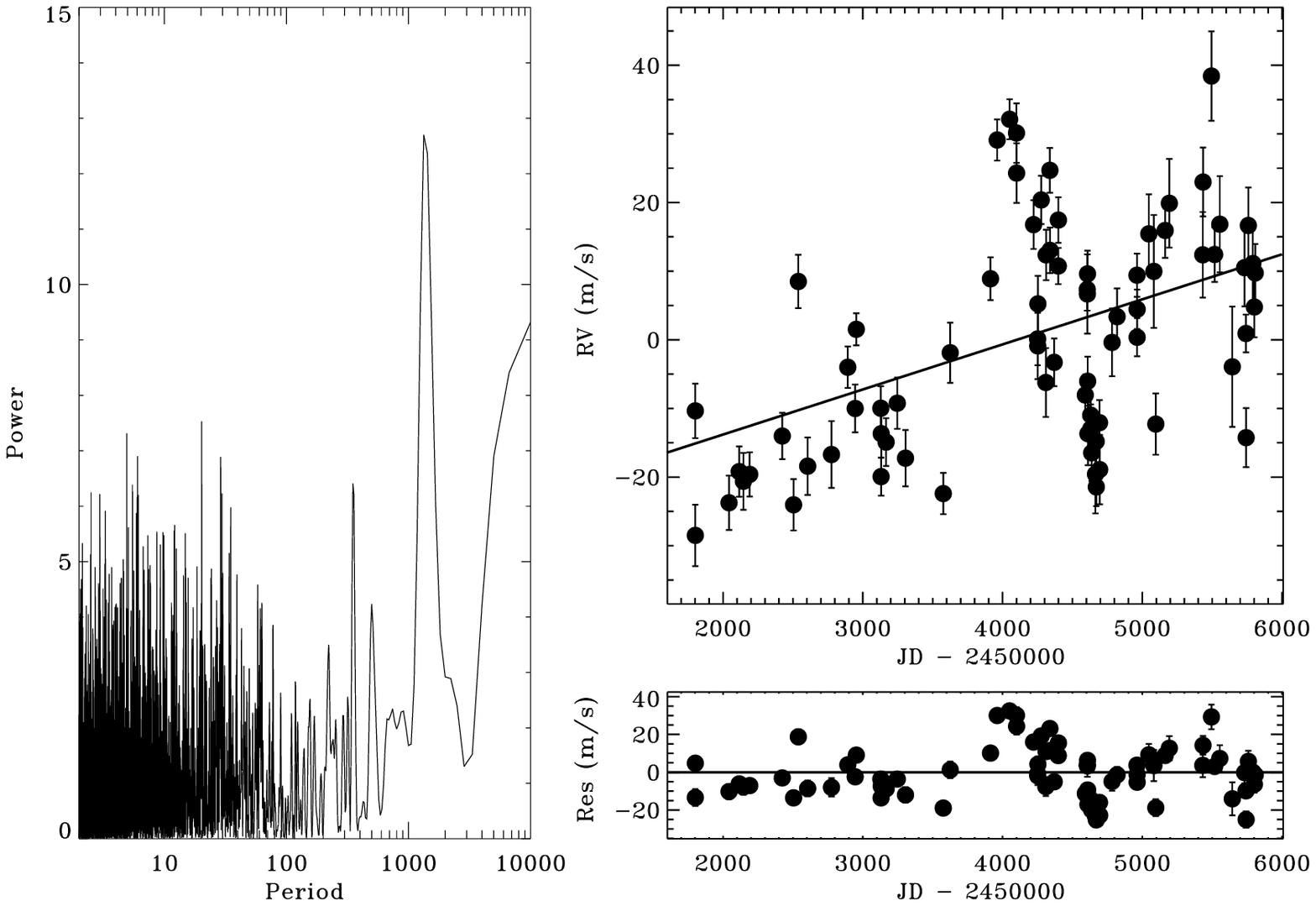}\\
\includegraphics[width=9cm]{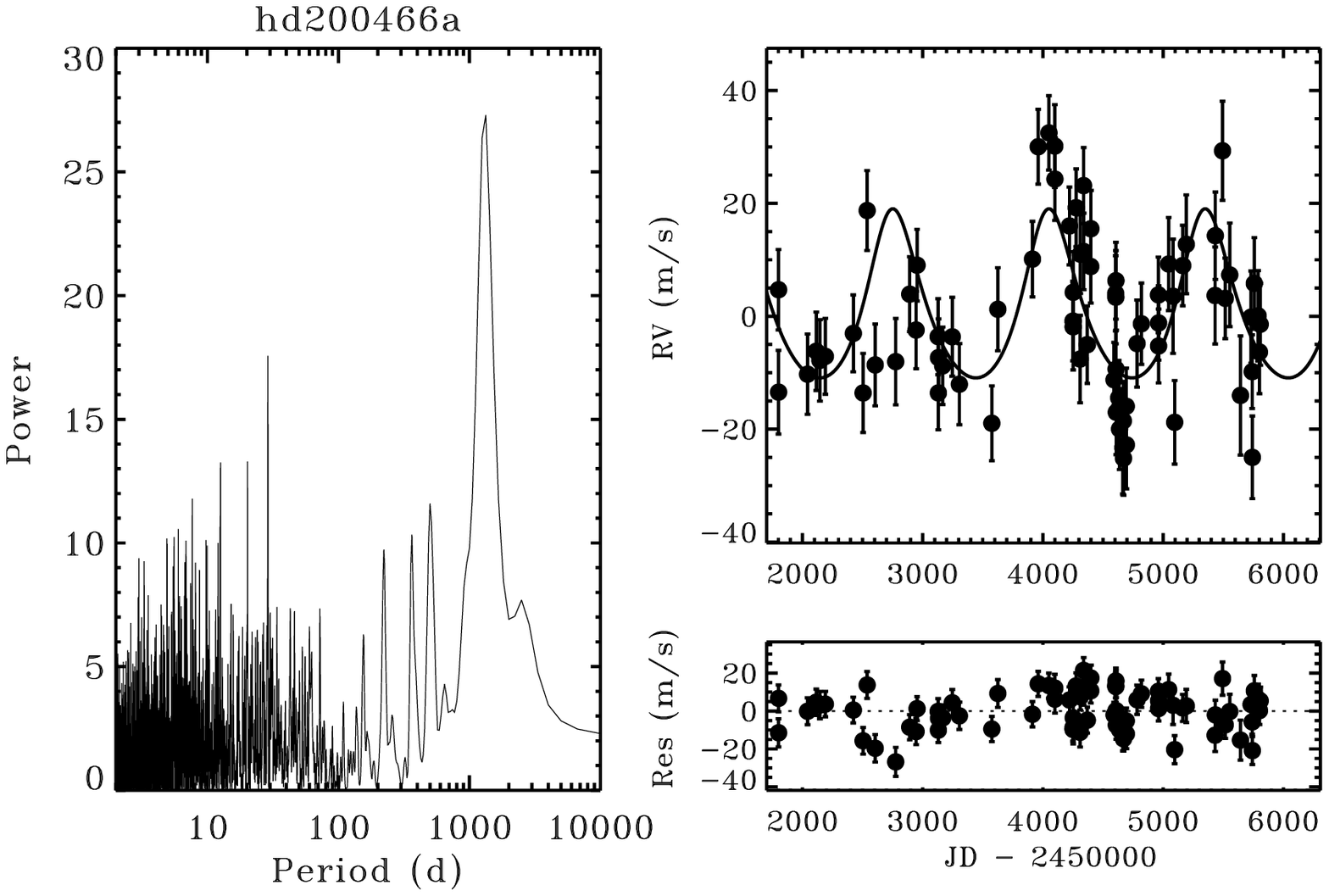}
\caption{Analysis of RVs for {\object HD~200466}A. Upper panels: measured RVs. 
Left panels: Lomb-Scargle periodogram. Right panels: temporal series of RVs and the 
residuals from the RV trend predicted by the binary orbital solution.
Lower panels: RVs corrected for binary motion. Left panel: Lomb-Scargle periodogram 
of the residuals. Right panels: RVs of the residuals. Overplotted is the best Keplerian 
fit of RVs including the RV trend derived from {\object HD~200466}B and the residuals (Sect.~6).} 
\label{f:rv_a}
\end{figure}

The source \object{{\object HD~200466}A} shows significant RV variability, more clearly seen
after JD 2453800 (Fig.~\ref{f:rv_a}). The RV scatter (16.0 m/s) is larger than that of 
the companion \object{{\object HD~200466}B} (10.8 m/s) and much larger than the 
internal errors (4.2 m/s). As our targets are moderately active, it is important to 
estimate the expected RV jitter. For stars with chromospheric activity and colours like 
those of {\object HD~200466} A and B, the calibration by Wright (2005) 
yields a median jitter of 5.9 and 7.0 m/s, with 20th percentiles of 3.9 and 4.5 m/s 
and 80th percentiles of 9.1 and 10.8 m/s for {\object HD~200466} A and B, respectively.
A RV jitter as high as 15 m/s, as that required to explain the RV scatter of 
{\object HD~200466}A, appears then very unlikely.

\begin{figure}
\includegraphics[width=9cm]{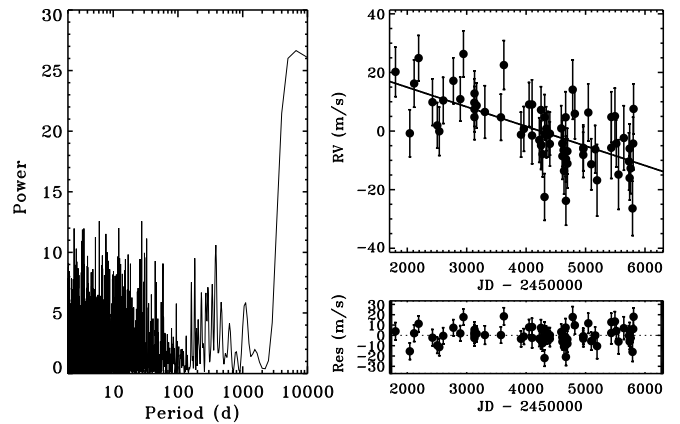}
\caption{Analysis of RVs for {\object HD~200466}B. Left panel: Lomb-Scargle 
periodogram. Upper right panel: temporal series of RVs. The RV trend predicted by the
binary orbital solution is overplotted.  Lower left panel: the residuals around this line}
\label{f:rv_b}
\end{figure}

We consider first the secondary. It shows a small (but clearly significant) downward 
trend ($-0.0068\pm0.0010$ m/s/d), with r.m.s. of RVs decreasing from 10.8 m/s to 8.3 m/s after 
removing the trend (Fig.~\ref{f:rv_b}). This latter can be fully explained by internal 
errors and our estimated activity jitter (7 m/s).


The primary has larger variations, with a long-period modulation that appears more 
clearly in the second half of the data and a possible general upward slope. 
Removing the slope yields only limited reduction of r.m.s. of the RVs (13.7 m/s)
The Lomb-Scargle periodogram of RVs shows a peak at about 1300~d, as well as additional 
power at longer periods (Fig.~\ref{f:rv_a}). A bootstrap test shows a false alarm 
probability of 0.3\% for the 1300 d peak for the original RV dataset.

We will show in Section~\ref{s:rvfit} that a single Keplerian orbit does not give 
a satisfactory fit of the RV curve. The residuals of one planet orbit show a clear 
long-term modulation (see Figure~\ref{f:rv_a}). Including a linear RV 
drift in the fit yields a slope very similar to that of {\object HD~200466}B with 
opposite sign ($+0.0071\pm0.0010$ m/s/d). The downward trend observed in HD~200466B 
and the possible upward trend in HD~200466A suggest the possibility that part of 
the observed RV variations are due to the binary motion of the wide pair. To further 
test this hypothesis we consider in Sect.~\ref{s:bin} the constraint we can put on the 
binary orbit.

\section[System architecture]{System architecture \label{s:bin}}

We studied the orbit of the wide binary with the goal of putting constraints on the 
system parameters. We discuss the astrometric data and the radial velocity difference
between the components.

\subsection{Astrometric data}
\label{s:astrometricdata}

\begin{figure}[h, t, b, p]
\includegraphics[width=9cm]{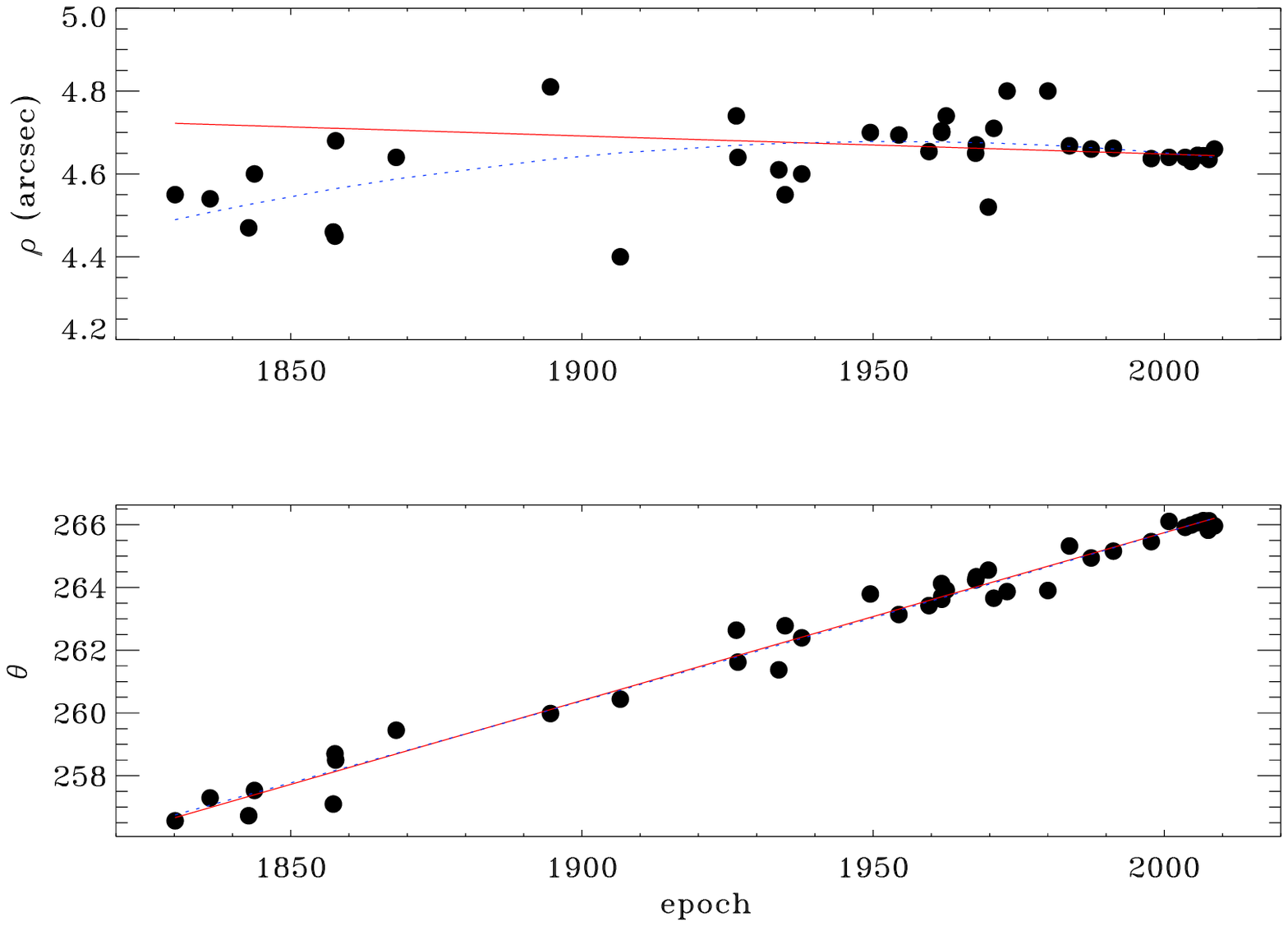}
\includegraphics[width=9cm]{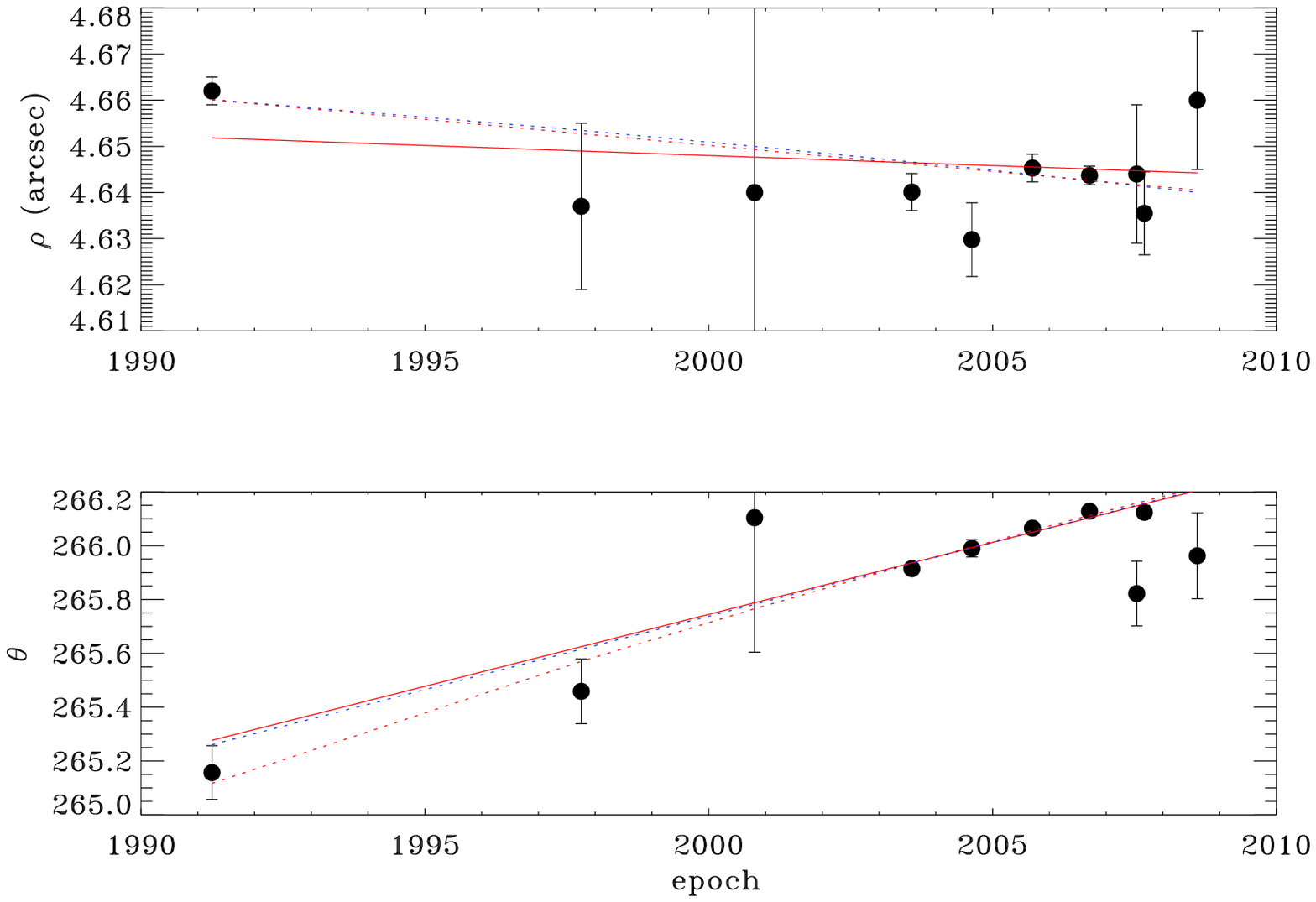}
\caption{\footnotesize Upper panel: 
Relative astrometry of {\object HD~200466}. From top to bottom: projected separation vs time.
Position angle (corrected for precession) vs time. Data from WDS (kindly 
provided by Dr. B. Mason) and from Table 2. Linear fits over the 
whole time extent of observations are shown as red solid lines. Blue dotted lines 
represent the quadratic fits and green dashed lines the linear fits made 
separately on data taken before and after 1950.0. The red dotted line is the fit 
on the recent high-quality measurements, extrapolated backward in time.
The lower panels displays the same quantities of the first two panels, plotting only the best data (Hipparcos, speckle
interferometry, adaptive optics). The red dotted line is the fit on the recent 
high-quality measurements, the red continuous line the linear fit over all the 
measurements, the blue dotted line the quadratic fit}
\label{f:rhotheta1}  
\end{figure}

Visual observations available in the Washington Double Star Catalog (WDS, Mason et 
al. 2001) were kindly provided by Dr.~B.~Mason (version 4 April 2011). They 
span the epoch from 1830 to 2007. Our AdOpt@TNG observations (Section~2.2) 
further extend the time baseline to 2008. Figure \ref{f:rhotheta1} shows the available 
measurements.

\begin{table}
\begin{center}
\caption{Relative astrometry of the components of {\object HD~200466} 
from the observations with AdOpt@TNG.}
\begin{tabular}{lcc}
\hline
\noalign{\smallskip}
Epoch & $\theta$ & $\rho$ \\
      & deg      & arcsec  \\
\noalign{\smallskip}
\hline
\noalign{\smallskip}
2007.528 & 265.88$\pm$0.12 & 4.643$\pm$0.015 \\
2007.550 & 265.69$\pm$0.12 & 4.645$\pm$0.015 \\
2008.605 & 265.92$\pm$0.16 & 4.660$\pm$0.015 \\
\noalign{\smallskip}
\hline
\end{tabular}
\end{center}
\label{t:adopt}
\end{table}

Orbital motion is clearly detected, with position angle changing from 257 to 
266~deg, while the projected separation changed very little. No change in position 
angle and projected separation are reported in the Hipparcos catalog. We provide error 
bars for $\rho$ and $\theta$ values through separate fitting data for different 
epochs and using the original references for some recent high-quality data (Hipparcos, 
speckle interferometry by Douglass et al. 1999, WDS catalog by Mason et 
al. 2001, CCD astrometry by Izmailov et al. 2010, and our AO data).
The projected separation increases up to epoch about 1960, with a shallow decreasing 
slope in the last years (better seen in lower panel of Figure \ref{f:rhotheta1}, 
which shows only recent high-quality data). To estimate the significance of the 
curvature we performed a simulation randomly generating a dataset of 10000 values following 
Gaussian distribution for each individual value of $\rho$ and $\theta$ and their error 
bars to study the change in slope of the parameters in different epochs by using 
linear and quadratic fits. No significant curvature of the position angle trend is 
present and a linear fitting yields a slope of $+0.0535\pm0.0010$ deg/yr for $\theta$.
For $\rho$, a slope equal to $-1.35\pm0.23$ mas/yr was derived using a 
quadratic fit at epoch 2007.0 that differs by more than $2 \sigma$ from that at epoch 
1850.0 ($+2.58\pm0.61$ mas/yr). This confirms a change in slope  of projected 
separation over time (slightly positive before 1950, slightly negative after).

There are two possible interpretations of the observed change in slope in projected 
separation:
\begin{enumerate}
\item It is due to the binary motion of the wide pair (and then can be used to 
constrain the orbit)
\item
It is due to an additional object with period of several decades (in this case, most 
likely the one which is responsible for the trend of RVs of {\object HD~200466}B)
\end{enumerate}

The second hypothesis was rejected after the stellar activity analysis and the 
orbital stability analysis (see Section \ref{s:rvfit}).

\subsection{RV difference between the components}
\label{s:rvdiff}

The spectra obtained with SARG allow the difference in radial velocity difference 
between the components to be determined with high precision. We obtain this 
difefrence by deriving the radial velocities of both components
using the template of {\object HD~200466}A. The similarity of the spectra of the two 
components ensures good quality of both RV components. The radial 
velocity difference $\Delta RV (B-A)$ is of $-436$ m/s (at epoch 2007.0).
Formal measurement errors are well below 10 m/s. True errors due to uncertainties 
on system velocity due to RV variability (see below), gravitational, and convective 
shifts, etc., likely exceed this value. We adopt 30 m/s as a conservative error bar on 
velocity difference. 

\subsection{Assuming RV trend of B due to binary orbital motion.}
\label{s:binwithtrend}

Having obtained position and velocity on the plane of the sky and the velocity along 
the line of sight, a family of possible bound orbits can be obtained as a function 
of the unknown separation along the line of sight $z$. We follow the approach by 
Hauser et al. (1999). When assuming that the curvature in projected separation is real and due 
to binary orbital motion, we considered position angle, projected separation, and 
their derivatives with time as resulting from the linear and quadratic fits (for 
$\theta$ and $\rho$, respectively) at the reference epoch 2007.0.

\begin{figure}
\includegraphics[width=9cm]{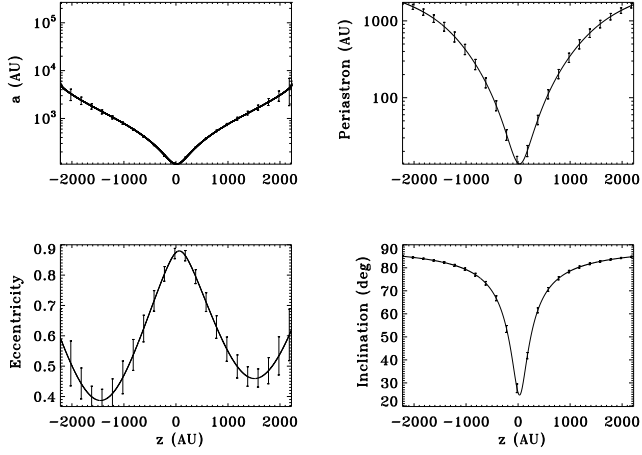}
\caption{Semi-major axis, periastron, eccentricity, and inclination of possible binary 
orbits as a function of the separation along the
line of sight $z$ (at reference epoch 2007.0), 
derived from relative positions and 
velocities on the plane of the sky and RV difference between the components. The probable curvature
in position angle vs time and the long-term RV trends are compatible with
a limited range in $z$. }   
\label{f:binorbit2}
\end{figure}

Figure \ref{f:binorbit2} shows the values of semi-major axis, eccentricity, periastron, 
and inclination for possible orbits. We obtained these values through a simulation of 
more than 1000 binary orbital parameter sets; we then calculated and plotted the curve 
that represents the mean values of random generations of the input parameters ($\rho$, 
$\theta$, $\delta\rho / \delta$t, $\delta\theta / \delta$t, $\Delta$RV, $\pi$, $m_{A}$, 
mass ratio) for each value of $z$ (with steps of $1$AU) with their $1 \sigma$ dispersion 
of simulated values at each distance. Either orbits with periastron close enough to 
induce significant perturbations to planets around the components or very wide orbits 
are possible.

\begin{figure}
\includegraphics[width=9cm]{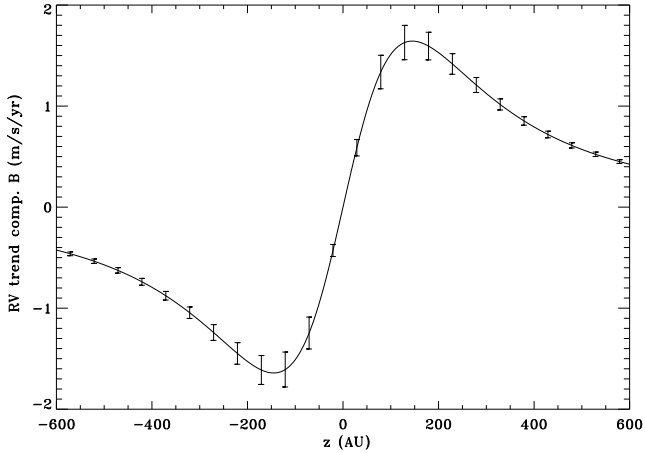}
\caption{Expected RV trend of B component at reference epoch 2007.0 as a function of $z$.}
\label{f:rvtrend}    
\end{figure}

\begin{figure}
\includegraphics[width=9cm]{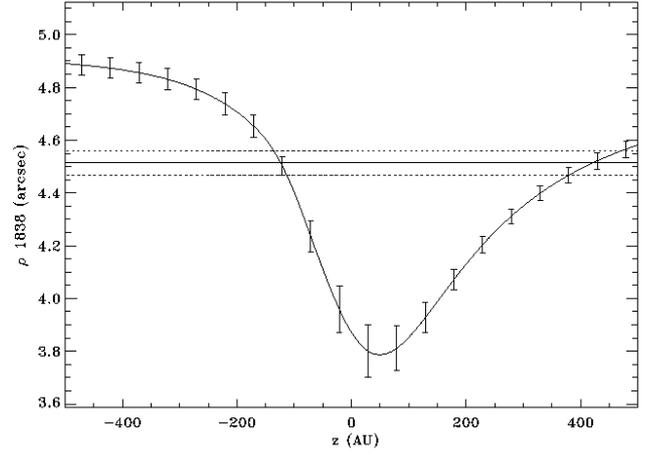}
\caption{Predicted projected separation and position angle at epoch 1840 as a
function of $z$. The measured values and the $1 \sigma$ uncertainties are 
overplotted as horizontal lines.}    
\label{f:rho1840}
\end{figure}

To further constrain the binary orbit we considered the change in slope in the 
projected separation described in Sect.~\ref{s:astrometricdata}. The first four measurements of projected 
separation have $\rho=4.540\pm0.027$~arcsec and position angle $\theta=257.03 
\pm 0.23$~deg (mean epoch 1838.21). Only a limited range of values of $z$ are 
compatible with these data (Fig.~\ref{f:rho1840}), with the projected separation providing the tightest 
constraints while the value of $\theta$\ at 1838 epoch is compatible with a broader range 
of $z$\footnote{These conclusions change only marginally if we consider a more 
extended data set to constrain the separation at early phases e.g. the first eight 
data points before epoch 1868.}.

Two groups of orbits are compatible with the astrometric data. They correspond 
to $z \sim -150$ and $z \sim 400$ AU. When considering the expected RV trends (for 
the B component), the $z \sim -150$ orbital solution is close to the absolute
minimum of the RV trend (Figure \ref{f:rvtrend}), which is only marginally lower than 
the observed one  ($-1.64\pm0.08$ vs $-2.48\pm0.37$ m/s/yr). Therefore, both the 
curvature in projected separation with time and the RV trend can be explained by 
the binary orbit and allow its full derivation.

\begin{table}
\begin{center}
\caption{Parameters of the binary orbit resulting from best fitting of available data 
(astrometry at 1840)and RV trend of {\object HD~200466}B}
\begin{tabular}{lc}
\hline
\noalign{\smallskip}
Parameter  & Value  \\
\noalign{\smallskip}
\hline
\noalign{\smallskip}
$z$ (AU)          & -123.5$\pm$23    \\
$a$ (AU)          &  130.7$\pm$7.0   \\
$P$ (yr)          &   1092$\pm$87    \\
$e$               &  0.842$\pm$0.008 \\
$i$ (deg)         &   41.9$\pm$3.2   \\
$\Omega$ (deg)    &   44.2$\pm$3.8   \\
$\omega$ (deg)    &  230.8$\pm$5.0   \\
$T0$ (Bess. Yr)   &   1469$\pm$22    \\
RVtrend A (m/s/yr)&  1.584$\pm$0.058 \\
RVtrend B (m/s/yr)& -1.614$\pm$0.058 \\
$a_{crit}$ A (AU) &   3.41$\pm$0.45  \\
$a_{crit}$ B (AU) &   3.37$\pm$0.45  \\
\noalign{\smallskip}
\hline
\end{tabular}
\end{center}
\label{t:binorbit}
\end{table}

\subsection{Assuming that $\rho$ curvature/RV trends are not due to binary orbit }
\label{s:nobin}

The alternative hypothesis is that {\object HD~200466}B has an additional
companion that is responsible for the observed RV trend. 

As we have seen in Sect.~\ref{s:binwithtrend}, if we adopt the slope in projected 
separation $\rho$ that results from the latest data, an orbital solution close to that
predicting a RV trend comparable to the observations does occur. In this section, 
assuming an additional companion is responsible for the RV trend of {\object 
HD~200466}B, we consider that the observed curvature in projected separation with 
time is due to astrometric variations with timescales of a few decades induced by 
the close companion of {\object HD~200466}B responsible for the RV trend. In this 
case, the binary  (computed using the secular projected separation trend) 
is marginally constrained\footnote{Only the values of $z$ close to 0 are 
excluded because they would imply a position angle different from that observed.}, 
with a wide family of possible orbits. 
Furthermore, some important parameters such as the RV difference between the components and
the individual masses can be somewhat altered by the companions, adding further 
uncertainties.


\section{First approach: a planet around A?}
\label{s:rvfit}

As discussed in Sect.~\ref{s:bin}, the most likely scenario to explain the RV, 
astrometric, and direct imaging data is that the RV trend of {\object HD~200466}B
is due to the orbital motion of the wide pair. We will adopt the binary orbit as 
derived following this hypothesis. In this case, the RVs of {\object HD~200466}A 
should be characterized by a similar slope with opposite sign to that of 
{\object HD~200466}B. To analyse the RVs we then removed the expected RV trend 
for component A, taking into account the small mass difference between the two
components. Figure~\ref{f:rv_a} shows the Lomb-Scargle periodogram of the RV 
corrected for binary motion. The 1300~d peak becomes more prominent and isolated.
The false alarm probability for that peak derived using the bootstrap technique is 1/10000.

Orbital fitting (after quadratically adding the estimated jitter of 7 m/s) was performed 
using an IDL code based on a Levenberg-Marquardt least-squares fit of RVs. The 
resulting orbital parameters are listed in Table \ref{t:fit200466}. We considered 
orbits derived after correcting the RVs for the calculated binary orbit (trend 1.6~m/s/yr;
Sect.~\ref{s:binwithtrend}) 
and those derived considering the opposite of the observed RV trend of {\object HD~200466}B 
(corrected for mass ratio between the components, 2.4~m/s/yr). The errors in the 
orbital parameters were derived simulating synthetic datasets taking into account the 
7~m/s jitter (quadratically summed to the RV error) and performing orbital 
fitting for each fake RV series.

Figure \ref{f:rv_a} shows that the orbital fitting is not fully satisfactory, 
with residuals larger than  the expected internal errors + jitter (r.m.s. 10.5 m/s), and with some coherent 
structure in the first 1500 days of our observations.
This calls for an in-depth study of the origin of the RV variations. 
In Section \ref{s:origrv} we will analyse the stellar activity and reveal what we
think is the true source of the RV variations.

\begin{table}
\caption{Orbital parameters and 
results of Keplerian fitting for RVs. In the 2nd column the parameters obtained removing 
the opposite of the trend measured for {\object HD~200466}B RVs are listed; 
in the 3rd column the parameters obtained removing the trend derived 
from the binary orbit (Table 3) are listed}
\begin{tabular}{lcc}
\hline
\noalign{\smallskip}
Parameter & Fit 1  & Fit 2 \\   
\noalign{\smallskip}
\hline
\noalign{\smallskip}
Period (d)    &   1286$\pm$26   &  1298$\pm$26   \\
K (m/s)       &   14.7$\pm$2.0  &  14.7$\pm$1.8  \\
e             &   0.18$\pm$0.10 &  0.18$\pm$0.10 \\
$\omega$      &    354$\pm$64   &   354$\pm$64   \\
T0            & 1472.0$\pm$151  &1452.7$\pm$155  \\
M$sini$       &   0.75$\pm$0.09 &  0.75$\pm$0.08 \\
a (AU)        &   2.27$\pm$0.03 &  2.29$\pm$0.03 \\
r.m.s. res (m/s) &  10.5           &  10.8          \\
red. $\chi^2$ &   1.33          &  1.38          \\
\noalign{\smallskip}
\hline
\end{tabular}
\label{t:fit200466}
\end{table}

The periodogram of the residuals shows the highest peak at 20.2 days, very 
close to the possible photometric period. 
The 20~d periodicity is more evident when considering only the data after 
JD 2453800. 
This is consistent with lack of coherence of rotational modulations over 
the whole baseline, due to finite lifetime of active regions.

The inclusion of a second planet with period about twice that of the 
first one might explain the unequal amplitude of the RV maxima. However, this additional 
planet should have a semi-major axis of about 3.6 AU, outside the critical value for 
orbital stability. Numerical integrations of the system orbit show that on a very short 
timescale the putative outer planet has an encounter with the inner one and the following 
evolution is chaotic. Therefore, we can safely neglect such a two--planet solution for 
the assumed binary orbit.

\section{The true origin of the RV variations}
\label{s:origrv}

To explain the reason for the unequal amplitude of the RV maxima we study the stellar 
activity and search for correlations between the RV and the BVS 
and measure an index on the $H_{\alpha}$ chromospheric emission. Activity induced RV 
variations or contamination by light from the other component are expected to produce 
such correlations (see e.g. Queloz et al. 2001; Martinez Fiorenzano et 
al. 2005), even if this diagnostic loses sensitivity for slowly rotating stars 
as the components of {\object HD~200466} (see e.g. Desort et al. 2007).

The few spectra taken in poor observing conditions (seeing about 2 arcsec) do not 
show signatures of contamination (as expected considering the projected separation of 
about 4.6 arcsec) and so they were kept in the analysis. Furthermore, as the 
spectra of the two components were usually taken close in time and with similar 
observing conditions, contamination would affect in a similar way the spectra of both 
components. The hypothesis of contamination is not considered further.

\begin{figure}
\includegraphics[width=9cm]{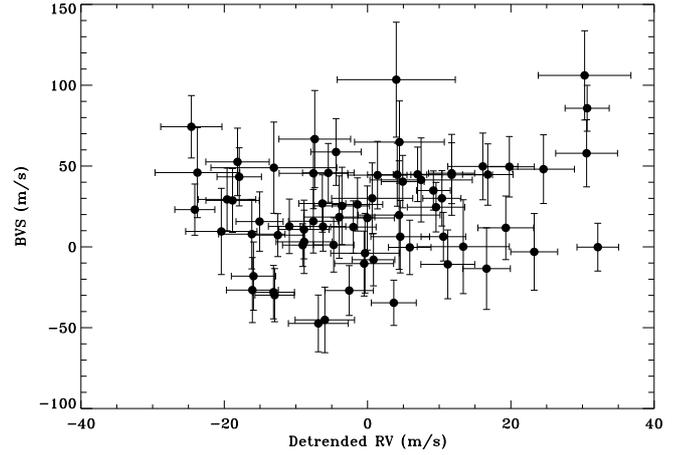}
\includegraphics[width=9cm]{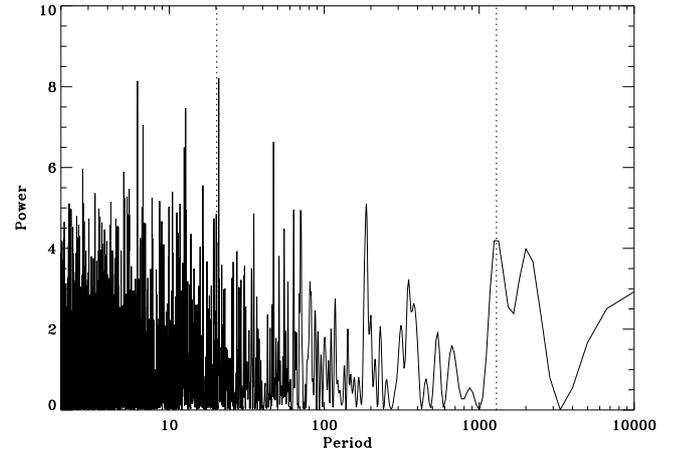}
\caption{Upper panel: BVS vs radial velocity (corrected for 
trend) for {\object HD~200466}A. Lower panel: Lomb-Scargle periodogram of BVS
of {\object HD~200466}A. The two vertical dotted lines mark the 
periods seen in the RVs time series (1300~d) and in the residuals from Keplerian orbit 
(20.25 days)}        
\label{f:bvsa}
\end{figure}

\begin{figure}
\includegraphics[width=9cm]{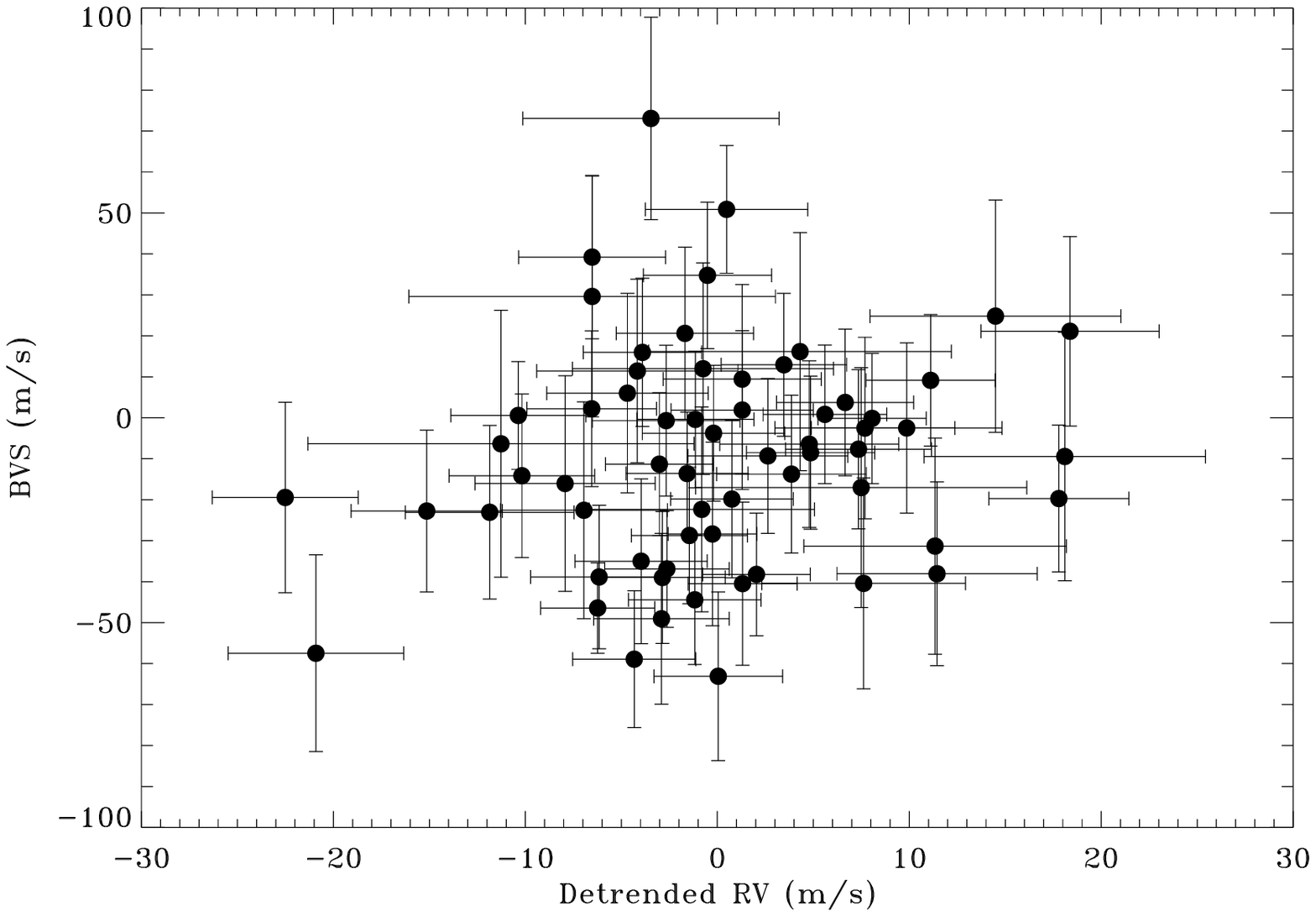}
\includegraphics[width=9cm]{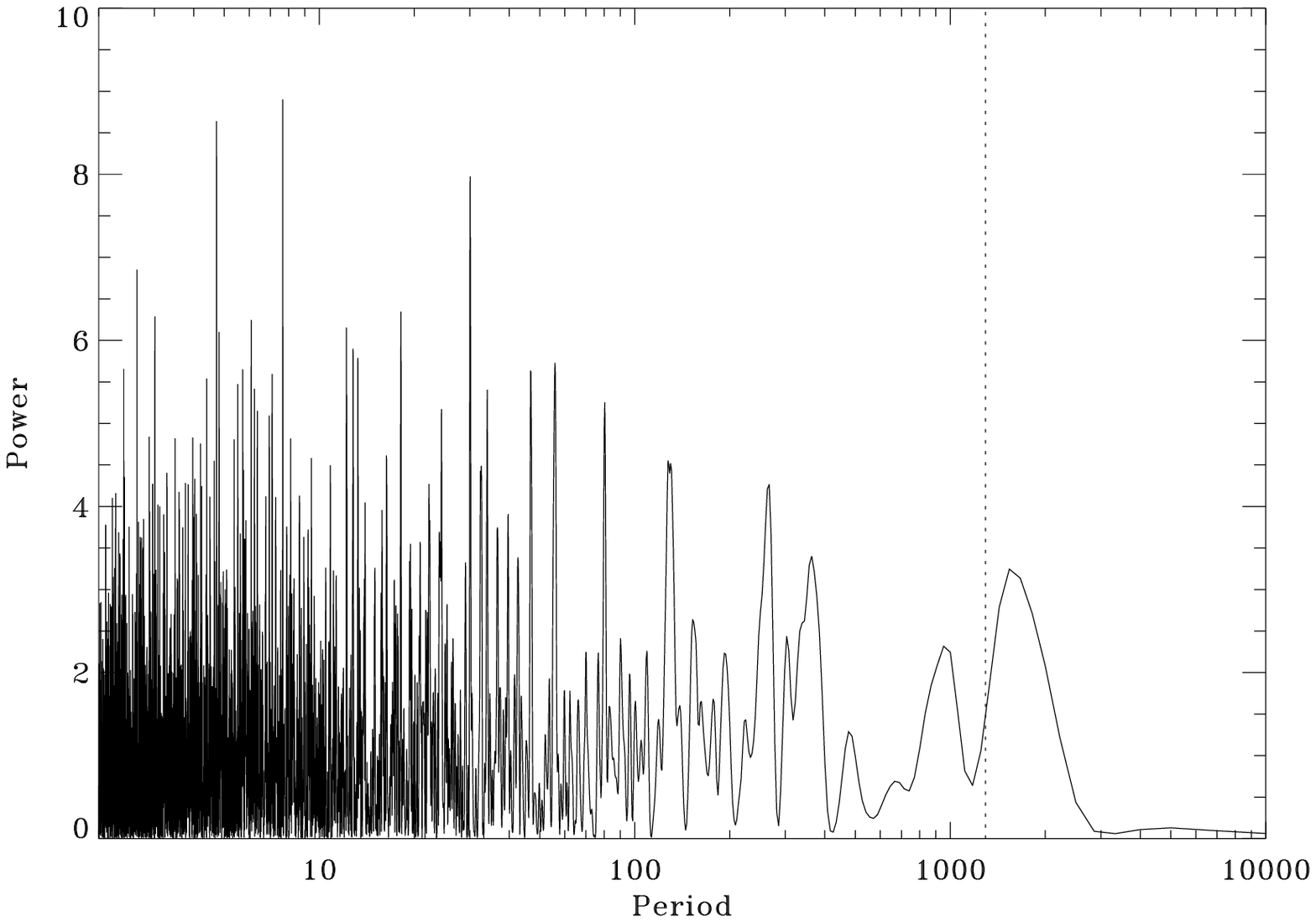}
\caption{ Upper panel: BVS vs radial velocity (corrected for 
trend) for {\object HD~200466}B. Lower panel: Lomb-Scargle periodogram of BVS
of {\object HD~200466}B. The vertical dotted line marks the 
period seen in the RVs time series of {\object HD~200466}A (1300~d)}        
\label{f:bvsb}
\end{figure}

Plots of RV vs BVS correlation of {\object HD~200466} A and B
are shown in Figs.~\ref{f:bvsa} and \ref{f:bvsb}, along with the Lomb-Scargle periodograms of the BVS;
this show a marginal correlation with the original (i.e. uncorrected for binary motion)
RVs of {\object HD~200466}A (Pearson correlation coefficient of 0.33 and Spearman rank correlation 0.25, with probability of 
occurring by chance of $<0.5$\ and 3.7\%, respectively). The correlation is weaker for the RVs corrected for binary 
motion (prob. 20\%). Lomb-Scargle periodograms show some power close to the RV period, 
but of low significance. One of the highest peaks at short periods is very close to the 
20~d periods seen in the Hipparcos photometric time series (Sect.~\ref{s:variability}) and 
RV residuals from Keplerian orbit (see Sect.~\ref{s:rvfit}). A similar analysis for the
B component yields a highest peak at about 30~d.
 
Of course, the 1300~d period is very different from the rotational period of the star,
which is most likely close to 20 days. It might instead be a magnetic activity 
cycle, characterized by a higher amplitude in the last years of our observations 
than in the first half.

\begin{figure}
\includegraphics[width=9cm]{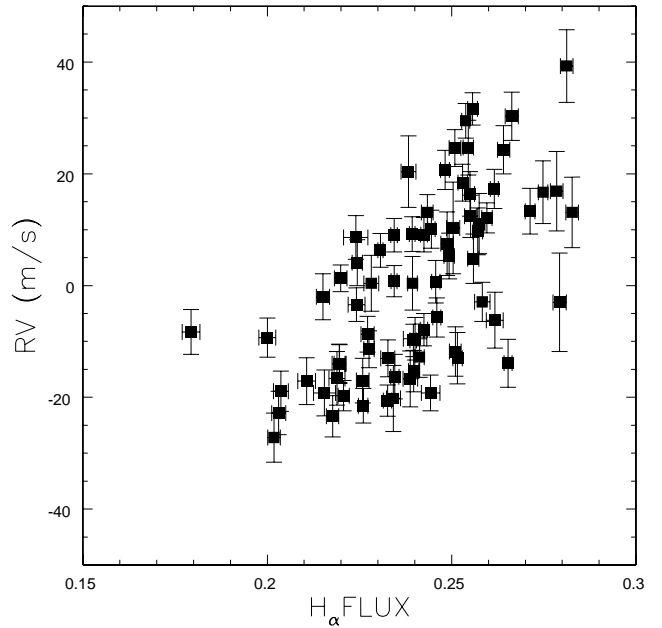}
\caption{RVs - $H_{\alpha}$ correlation for {\object HD~200466}A.}     
\label{f:halpharelationAa}
\end{figure}

\begin{figure}
\includegraphics[width=9cm]{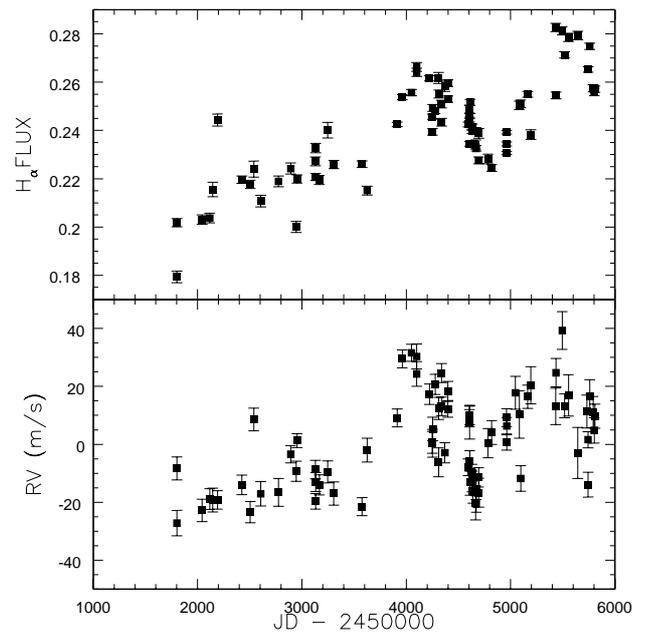}
\caption{$H_{\alpha}$ vs JD and RVs time series for {\object HD~200466}A. The modulation 
of the RV variations is due to the stellar activity cycle.}     
\label{f:halpharelationAbc}   
\end{figure}

\begin{figure}
\includegraphics[width=9cm]{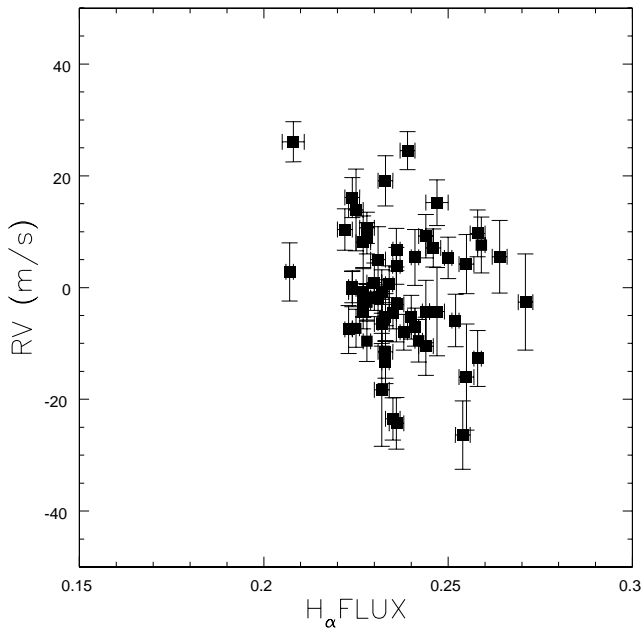}
\caption{{\object HD~200466}B: RVs - $H_{\alpha}$ correlation.}      
\label{f:halpharelationB}    
\end{figure}

In Figure \ref{f:halpharelationAa} the $H_\alpha$ index of {\object HD~200466}A 
correlates with the RV, indicating that the stellar activity could be responsible 
for the RVs variations. In particular in Figure \ref{f:halpharelationAbc} we show 
the time dependence of the activity index compared with the RV time series. The two 
plots look very similar to each other. This is a clear signature of the large variation 
of the activity of {\object HD~200466}A. We also measured the activity index for 
{\object HD~200466}B (see Fig.~\ref{f:halpharelationB}); in this case there is no 
signature at long periods; the strongest peak in the periodogram is at 12.7~d. 
However, the S/N of this detection is low (2.7), so that this result has a low confidence level.

In summary, although there is not a highly significant correlation between RVs and line 
bisectors, there is a clear indication of activity from the correlation between RVs and 
$H_\alpha$ line core strength. We argue that the RV variations observed in the primary 
component is related to the stellar activity cycle.

We also note that the relation between rotational period and length of the activity cycle has been 
discussed by several authors (see e.g. Baliunas et al. 1996; Olah et al. 2009) because it provides 
useful information on the dynamo mechanism (Durney \& Stenflo 1972). If we compare our results
for {\object HD~200466}A with those for other stars (see Figure 5 of Olah et al. 2009), we find 
that this star falls in a well populated region of this diagram close to the relation for ``shortest 
cycle lengths"; a similar result is obtained considering data from Lovis et al. (2011). In addition, 
a long-term trend is clearly present in our activity measures. It suggests that additional longer 
cycles should be present, which is a common finding for solar-type stars (see Baliunas et al. 1996, 
and Olah et al. 2009). On the other hand, the lack of a detectable cycle in {\object HD~200466}B
is not at all anomalous; Lovis et al. (2011) detected cycles only in about a third of 
the FGK stars with activity levels similar to those of {\object HD~200466}A and B.

\section{A corrected radial velocity series for HD200466A}
\label{s:correction}

Lovis et al. (2011) proposed two different approaches to correct radial velocities for the 
signal due to the activity cycle, and Dumusque et al. (2011) showed how the signal due to
planets can be extracted when such corrections are included in the data analysis.
We explored such approaches for the case of HD200466A. The basic idea is that long-term
activity should be positively correlated with radial velocities, essentially because magnetic
fields related to activity inhibit convection and then reduce the blueshift of spectral lines 
due to the ascending hot bubbles in the subatmospheric convection zones. At a first approximation, 
we expect that the radial velocity signal should then be corrected for a factor which is
proportional to the activity level, as measured by the emission in the H$_\alpha$\ core.
The conversion factor may be derived empirically by fitting the radial velocity curve vs H$_\alpha$. 
There is only one free parameter to be determined. 

A periodogram analysis of the time series for the H$_\alpha$\ index gives a period of $1413\pm 38$~d
and an amplitude of $0.0155\pm 0.014$. Following Dumusque et al. (2011), we then fitted the radial velocity 
curve for HD200466A with a sinusoid having the same period and phase of the solution found for 
the H$_\alpha$\ index, after having subtracted a linear trend corresponding to the binary orbital 
motion. The best fit is obtained for an amplitude of $10.5$~m/s. Subtraction of this signal 
reduces the r.m.s. scatter of the radial velocities from 15.9 m/s down to 11.6 m/s.

A periodogram analysis of these corrected radial velocities does not yield a single dominant
peak. The strongest peak is at a period of $20.241\pm 0.012$~d with an amplitude of 
$8.3\pm 1.6$~m/s; it coincides with the photometric signal obtained from the Hipparcos data 
(see Section~\ref{s:photom}) that is also present in the periodogram of the bisectors (see 
Section~\ref{s:origrv}) and can be interpreted as the rotational period of the star. Inclusion 
of this signal reduces the r.m.s. scatter of the radial velocities down to 10.2 m/s.

Once a sinusoid corresponding to this peak is subtracted, the three next highest peaks are at 
periods of $27.918\pm 0.014$~d, $1102.24\pm 0.016$~d, and $2702\pm 30$~d. While we deem the rotational 
period to be quite robust, because it is present in completely independent time series, although never at 
a high level of significance, the remaining peaks are not robust; their frequencies change if, 
for instance, we drop the last four points of the time series.

A similar approach has been criticized by Meunier \& Lagrange (2013) because it neglects the
fact that activity and radial velocity variations may be not sinusoidal with time. They then suggested
adopting a linear relation between the activity indicator (in our case, the H$\alpha$ index) and
the radial velocities, and then removing such a trend.
We applied a correction obtained by averaging the two results obtained using radial velocity and 
H$\alpha$\ index as independent variables. The corrected radial velocities have a very strong trend with time, with a slope of
-0.0117 m/s/day. This is a consequence of the presence of a quite strong trend of the H$\alpha$\ index
with time, while radial velocities only have a more modest trend. A Fourier analysis of
the residuals along this linear trend has the three highest peaks at 1145, 5.47, and 19.36~d. 
The first peak resembles the long period that we attribute to activity, possibly signalling that
the correction to be applied is more complex than a simple linear function. The third period
is possibly related to the power at about 20~d that we attribute to rotation.
 
\section{Discussion and conclusion}
\label{s:discussion}

\subsection{Lithium difference between stars with and without planets?}
\label{s:li_disc}

It is well know that lithium abundances in old solar-type stars show a large scatter, 
both in clusters and field (e.g. Pasquini et al. 1997 for the open cluster
M67). Large differences were also reported in a few cases of wide binaries with similar 
components (King et al. 1997 and Martin et al. 2002). The star {\object HD~200466} 
is a rather extreme case in this context, because of the large difference in lithium 
levels between the components (by a factor of at least 9) and because the 
difference is in the opposite direction with respect to the usual pattern of lower
lithium content toward lower temperatures, and therefore should have a different origin.
The very small errors in the temperature difference ($53\pm 23$~K) makes this result very robust.

In the past few years, there were several attempts to correlate such variations with the 
presence of planets, with indications that stars with planets typically have lower lithium 
abundance while some other works reached the opposite conclusion (see Gonzalez 2006 
for a summary and references). The work by Israelian et al.~(2009) supports the 
reality of such a correlation and theoretical models have been developed to explain these 
results. It is possible that a slow stellar rotation resulting from a longer star-disk 
interaction phase affects the core-envelope coupling. This might produce a stronger 
differential rotation at the base of the convective envelope and then larger lithium 
depletion in slow rotators (Bouvier 2008). Castro et al.~(2009) 
considered instead the possible role of angular momentum transfer due to planetary migration.

The components of {\object HD~200466} are slightly cooler than the limits of the sample
considered in Israelian et al.~(2009). We do not have an indication of the 
presence of giant planets within a few AU.
The difference in the activity cycle is the other peculiar feature of {\object HD~200466}, 
so one may think about a possible link between Li abundances and stellar activity. 
However, a comparison of the Li abundances (from
Sousa et al. 2010) in stars with and without activity cycles (from Lovis et al. 2011) does not
show any correlation between these two quantities, likely because activity is variable on a 
large range of timescales and detection of cycles in given stars might be related to
the epoch of observation. A systematic investigation of the characteristics of 
the activity cycles in multiple systems with similar components and the possible correlations 
of the activity parameters with other stellar properties is postponed to a forthcoming work.

\subsection[Conclusion]{Conclusion}
\label{s:li_disc}

The long-term RV trends of both components have opposite slopes and similar amplitudes. This 
fact suggests that we are observing the binary motion, and the available astrometric 
measurements as well as the non-detection of additional close companions in our adaptive 
optics images support this scenario. Strong constraints are put on the orbital solution of 
the binary, which is very eccentric and with a periastron as small as 22~AU, limiting 
the zone for dynamical stability around {\object HD~200466} to slightly more than 3 AU.

\begin{figure}
\includegraphics[width=8cm]{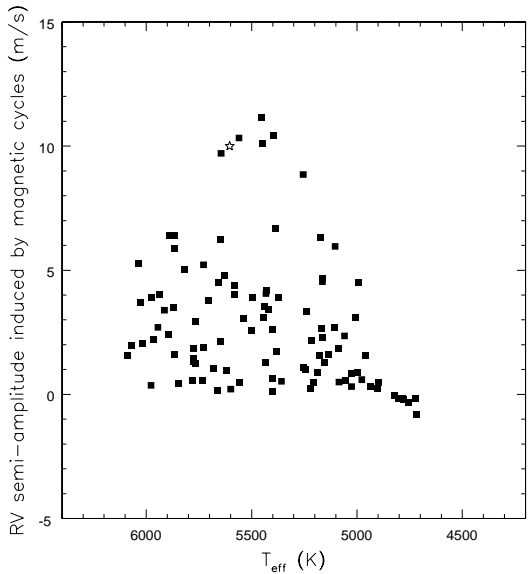}      
\caption{Adapted from Lovis et al. (2011). Effective impact of magnetic 
cycles on RV for HARPS sample stars with a detected cycle, shown as a function of $T_{eff}$. 
The location of {\object HD~200466}A in this diagram is shown as an open star.} 
\label{fig:lovis11}
\end{figure}

A vigorous activity cycle is present in {\object HD~200466}A during the second half of 
the observing range. This phenomenon mimics the 
presence of a planetary system with a period of 1300~d. As the star is moderately young 
(about 2~Gyr) and active (expected RV jitter about 5.9~m/s), the possibility of activity-induced 
jitter was considered. We did not detect a strong correlation between RV and BVS, while 
the $H_\alpha$ analysis leads a different conclusion. The star {\object HD~200466}A shows a
cycle amplitude as high as the maximum ones seen in the HARPS sample (see Fig.~\ref{fig:lovis11} 
adapted from Lovis et al.~2011). The jitter value is higher than the one 
predicted by Wright (2005) for {\object HD~200466}A (5.9 m/s). 

The residuals from a sinusoidal fit with period and phase set by the $H_\alpha$ observations
show the highest peak at about 20~d, very close to the probable (FAP 2.5\%) photometric period 
derived from the Hipparcos photometry. This may be interpreted as the rotational period of
the star.

We conclude this paper by stressing the importance of distinguishing between the activity cycles 
and planetary signatures. We studied the stellar activity of {\object HD~200466}A and 
estimated that the cycle semi-amplitude could amount to levels up to $>10$ m/s, with
rather large variations from one cycle to another. Therefore,
the monitoring of activity indicators is necessary in order to solve the true origin of the 
RV variations, whether they are due to a planet or to the stellar activity.

\begin{acknowledgements}

This research has made use of the SIMBAD database, operated at CDS, Strasbourg, France.
This research has made use of the Washington Double Star Catalog maintained at the U.S. Naval 
Observatory. This research has made use of the Keck Observatory Archive (KOA), which is 
operated by the W. M. Keck Observatory and the NASA Exoplanet Science Institute (NExScI), 
under contract with the National Aeronautics and Space Administration. We thank the TNG staff 
for contributing to the observations and the TNG TAC for the generous allocation of observing 
time. We thank R.~Ragazzoni and A.~Ghedina for useful discussions on AdOpt@TNG. We thank 
B. Mason for providing the astrometric data collected in the Washington Double Star Catalog.
This work was partially funded by PRIN-INAF 2008 ``Environmental effects in the formation and 
evolution of extrasolar planetary systems''. E.C. acknowledges support from CARIPARO.
We warmly thank the anonymous referee for her/his careful reading and detailed comments to
the manuscript.

\end{acknowledgements}

\Online

\begin{table*}[h]
\caption{Differential radial velocities of {\object HD~200466}A}
\label{t:rva}
\centering
\begin{tabular}{ccccccc}
\hline
\noalign{\smallskip}
JD - 2450000  &  RV & Error & BVS & Error & H$\alpha$ & Error\\
              & m/s & m/s   & m/s & m/s  & & \\
\noalign{\smallskip}
\hline
\noalign{\smallskip}
1801.5946 &  -8.3 & 4.0 & -57.4 & 17.5 & 0.1793 & 0.0024 \\
1801.6120 & -27.2 & 4.4 &       &      & 0.2018 & 0.0017 \\
2042.6776 & -22.8 & 3.9 &   3.1 & 19.6 & 0.2031 & 0.0019 \\
2115.6340 & -18.9 & 3.6 &  45.8 & 18.2 & 0.2037 & 0.0020 \\
2145.4529 & -19.2 & 4.1 & -45.2 & 20.3 & 0.2155 & 0.0030 \\
2190.5009 & -19.2 & 3.2 &  12.6 & 15.2 & 0.2443 & 0.0025 \\ 
2423.6701 & -14.0 & 3.4 & -27.0 & 15.4 & 0.2196 & 0.0015 \\
2504.6172 & -23.4 & 3.7 &   7.3 & 13.3 & 0.2177 & 0.0017 \\
2538.3752 &   8.6 & 3.9 &  11.8 & 19.7 & 0.2240 & 0.0033 \\
2605.3166 & -17.1 & 4.2 & -47.4 & 17.6 & 0.2107 & 0.0024 \\
2775.7122 & -16.6 & 4.8 &  45.5 & 21.9 & 0.2189 & 0.0022 \\
2892.4668 &  -3.4 & 3.0 &  40.4 & 16.2 & 0.2242 & 0.0023 \\
2945.3949 &  -9.3 & 3.5 &  26.2 & 16.5 & 0.2000 & 0.0023 \\
2953.4334 &   1.3 & 2.4 &  34.9 & 12.1 & 0.2199 & 0.0017 \\
3128.7347 &  -8.7 & 3.2 &  12.2 & 16.2 & 0.2272 & 0.0018 \\
3130.7155 & -19.7 & 2.7 & -29.9 & 16.4 & 0.2207 & 0.0015 \\
3131.7109 & -13.0 & 3.3 &  26.8 & 17.7 & 0.2327 & 0.0019 \\
3166.6250 & -14.0 & 3.5 &  15.8 & 19.7 & 0.2195 & 0.0019 \\
3246.4515 &  -9.5 & 3.8 &  25.3 & 23.9 & 0.2401 & 0.0032 \\
3305.3871 & -17.0 & 4.0 & -18.1 & 21.1 & 0.2259 & 0.0017 \\
3575.5412 & -21.5 & 3.1 &  43.3 & 18.0 & 0.2260 & 0.0013 \\
3625.3901 &  -2.0 & 4.1 &  44.4 & 20.9 & 0.2151 & 0.0018 \\
3913.5286 &   9.1 & 3.1 &   6.3 & 15.0 & 0.2425 & 0.0009 \\
3961.4901 &  29.5 & 3.1 &  85.7 & 14.2 & 0.2538 & 0.0011 \\
4050.3779 &  31.6 & 2.9 &  -0.2 & 14.8 & 0.2557 & 0.0013 \\
4099.3823 &  30.3 & 4.3 &  57.9 & 20.8 & 0.2663 & 0.0018 \\
4100.3198 &  24.3 & 4.3 &  48.1 & 21.3 & 0.2641 & 0.0017 \\
4221.6957 &  17.3 & 3.5 &  44.7 & 19.0 & 0.2616 & 0.0011 \\
4250.6626 &   0.7 & 3.8 &  17.9 & 19.8 & 0.2457 & 0.0011 \\
4251.6260 &   0.4 & 4.8 &  -3.9 & 24.7 & 0.2394 & 0.0013 \\
4252.6646 &   5.3 & 4.1 &   6.3 & 22.4 & 0.2493 & 0.0014 \\
4276.5873 &  20.7 & 3.5 &  49.6 & 18.7 & 0.2482 & 0.0012 \\
4309.5807 &  -6.2 & 5.0 &  66.7 & 30.0 & 0.2617 & 0.0023 \\
4311.5436 &  12.4 & 3.8 & -10.8 & 21.3 & 0.2551 & 0.0016 \\
4338.4765 &  24.6 & 3.3 &  -3.1 & 23.7 & 0.2509 & 0.0016 \\
4339.4697 &  13.1 & 3.2 &  45.6 & 18.8 & 0.2434 & 0.0016 \\
4369.5190 &  -2.9 & 3.5 &  58.7 & 20.7 & 0.2583 & 0.0021 \\
4398.3785 &  12.1 & 2.7 &  30.0 & 17.1 & 0.2596 & 0.0014 \\
4399.3194 &  18.4 & 3.3 & -13.5 & 25.2 & 0.2530 & 0.0013 \\
4591.7004 &  -7.9 & 2.9 &  12.7 & 16.9 & 0.2425 & 0.0010 \\
4605.6724 &   9.0 & 3.0 &  -0.3 & 16.8 & 0.2344 & 0.0009 \\
4606.6758 &   7.5 & 5.7 &  19.7 & 33.6 & 0.2488 & 0.0017 \\
4607.7101 &  10.1 & 3.4 &  44.9 & 16.9 & 0.2445 & 0.0010 \\
4608.6989 &  -5.7 & 3.5 &  10.7 & 18.2 & 0.2460 & 0.0011 \\
4610.6951 & -13.0 & 4.6 &   7.9 & 21.5 & 0.2516 & 0.0013 \\
4630.6642 &  -9.8 & 2.9 & -28.0 & 16.5 & 0.2396 & 0.0009 \\
4631.6686 & -12.8 & 3.6 & -26.7 & 20.1 & 0.2413 & 0.0013 \\
4637.6541 & -16.3 & 4.0 &  29.4 & 19.4 & 0.2346 & 0.0011 \\
4664.6267 & -20.2 & 5.9 &  45.9 & 27.8 & 0.2342 & 0.0019 \\
4669.6091 & -15.3 & 3.7 &  28.7 & 19.8 & 0.2397 & 0.0014 \\
4670.6013 & -20.6 & 2.8 &  23.0 & 15.8 & 0.2326 & 0.0011 \\
4693.6243 & -11.4 & 3.3 &  15.7 & 18.4 & 0.2276 & 0.0014 \\
4694.6057 & -16.7 & 5.0 &   9.5 & 26.7 & 0.2388 & 0.0020 \\
4783.4144 &   0.4 & 5.0 &  18.4 & 25.5 & 0.2282 & 0.0020 \\
4819.3206 &   4.1 & 4.1 & -10.4 & 20.1 & 0.2244 & 0.0014 \\
4961.6250 &   9.2 & 3.1 & -34.6 & 14.0 & 0.2393 & 0.0008 \\
4962.6310 &   0.8 & 2.8 &  -8.0 & 16.2 & 0.2344 & 0.0008 \\ 
4962.6570 &   6.3 & 3.0 &       &      & 0.2307 & 0.0007 \\
5046.4571 &  17.8 & 5.7 &  44.6 & 25.0 &        &        \\
5082.6078 &  10.3 & 8.2 & 103.4 & 35.5 & 0.2504 & 0.0018 \\
5096.4713 & -11.8 & 4.4 &  52.6 & 20.9 & 0.2510 & 0.0016 \\
5163.4122 &  16.4 & 4.0 &  24.5 & 15.3 & 0.2550 & 0.0012 \\
5193.3399 &  20.4 & 6.4 &   0.1 & 29.0 & 0.2383 & 0.0020 \\
5432.4439 &  13.1 & 6.3 &  64.8 & 25.4 & 0.2827 & 0.0017 \\
5434.4214 &  24.7 & 4.9 &  49.8 & 20.6 & 0.2545 & 0.0014 \\
5494.4058 &  39.3 & 6.5 & 106.1 & 27.6 & 0.2812 & 0.0017 \\
5517.3132 &  13.3 & 4.1 &  44.6 & 19.5 & 0.2713 & 0.0013 \\
5554.3615 &  16.9 & 7.1 &  41.4 & 26.0 & 0.2785 & 0.0017 \\
5643.7324 &  -3.0 & 8.8 &  49.0 & 28.2 & 0.2793 & 0.0018 \\ 
5731.5882 &  11.3 & 5.7 &  30.0 & 21.9 &        &        \\
5741.6275 &   1.6 & 2.8 &   1.1 & 13.2 &        &        \\
5742.7212 & -13.9 & 4.3 &  74.3 & 19.3 & 0.2653 & 0.0011 \\
5758.5279 &  16.7 & 5.6 &       &      & 0.2748 & 0.0013 \\
5790.5474 &  11.1 & 5.4 &       &      & 0.2575 & 0.0015 \\
5801.5488 &   4.8 & 4.4 &       &      & 0.2559 & 0.0015 \\ 
5807.6077 &   9.7 & 4.2 &       &      & 0.2571 & 0.0013 \\
\noalign{\smallskip}
\hline
\end{tabular}
\end{table*}

\begin{table*}[h]
\caption{Differential radial velocities of {\object HD~200466}B}
\label{t:rvb}
\begin{tabular}{ccccccc}
\hline
\noalign{\smallskip}
JD - 2450000  &  RV & Error & BVS & Error & H$\alpha$ & Error\\
              & m/s & m/s   & m/s & m/s  & & \\
\noalign{\smallskip}
\hline
\noalign{\smallskip}
 1801.5922 &  19.1 & 4.5 & -87.7 & 17.5 & 0.233  & 0.002  \\
 2042.6909 &  -0.7 & 3.9 & -22.8 & 19.7 & 0.232  & 0.002  \\
 2115.6491 &  15.2 & 4.1 &   9.4 & 23.1 & 0.247  & 0.003  \\
 2190.5138 &  24.5 & 3.4 &   9.1 & 16.1 & 0.239  & 0.002  \\
 2423.6818 &   9.2 & 3.9 &  -0.7 & 18.3 & 0.244  & 0.002  \\
 2504.6294 &   0.9 & 3.5 &   0.6 & 13.2 & 0.230  & 0.002  \\
 2538.3896 &  -0.8 & 4.4 & -23.1 & 21.2 & 0.227  & 0.004  \\
 2605.3300 &  10.4 & 3.7 &  -3.8 & 16.6 & 0.222  & 0.002  \\
 2775.7244 &  16.1 & 3.6 &   3.7 & 17.9 & 0.224  & 0.002  \\
 2892.4783 &  10.7 & 2.8 & -38.2 & 15.0 & 0.228  & 0.002  \\
 2945.4073 &  26.1 & 3.6 & -19.7 & 17.9 & 0.208  & 0.003  \\
 3128.7469 &   9.7 & 4.2 &  -9.3 & 18.9 & 0.258  & 0.002  \\
 3129.7328 &   4.2 & 3.3 & -39.0 & 16.1 &        &        \\
 3130.7033 &   7.1 & 3.4 & -63.1 & 20.6 & 0.246  & 0.002  \\
 3131.7234 &  12.6 & 3.2 &   0.8 & 16.9 &        &        \\
 3166.6379 &   7.6 & 3.2 & -19.8 & 19.2 &        &        \\
 3305.3991 &   5.0 & 5.9 & -22.4 & 25.0 & 0.231  & 0.002  \\
 3575.5529 &   5.3 & 3.7 &   1.9 & 19.4 & 0.250  & 0.001  \\
 3625.4015 &  22.1 & 4.6 &  21.1 & 23.1 &        &        \\
 3913.5398 &  -2.6 & 3.2 & -58.9 & 16.7 & 0.228  & 0.001  \\
 3961.5016 &  -0.2 & 3.2 & -13.6 & 15.0 & 0.224  & 0.001  \\
 4050.3894 &   8.8 & 2.8 &  -0.1 & 15.9 & 0.228  & 0.001  \\
 4099.3955 &  -3.0 & 6.7 &  73.1 & 24.7 & 0.236  & 0.002  \\
 4100.3320 &   8.1 & 4.7 &  -2.5 & 22.2 & 0.227  & 0.002  \\
 4221.7074 &  -4.3 & 3.1 &  16.0 & 18.1 & 0.227  & 0.001  \\
 4250.6743 &   6.8 & 3.8 &  -7.7 & 19.4 & 0.236  & 0.001  \\
 4251.6382 &  -7.5 & 4.3 & -22.6 & 26.5 & 0.223  & 0.001  \\
 4252.6759 &  -5.3 & 4.2 &   6.0 & 24.3 & 0.233  & 0.001  \\
 4276.5986 &  -7.3 & 3.4 &   2.2 & 19.0 & 0.225  & 0.001  \\
 4309.5925 & -23.5 & 3.8 & -19.5 & 23.3 & 0.235  & 0.002  \\
 4311.5553 &   3.9 & 3.3 &  -8.5 & 18.7 & 0.236  & 0.001  \\
 4338.4883 &   0.1 & 2.8 & -40.5 & 19.9 & 0.224  & 0.001  \\
 4339.4824 &  -2.3 & 3.0 &  -0.4 & 16.7 & 0.227  & 0.001  \\
 4369.5312 &  -1.9 & 3.3 &  34.8 & 17.9 & 0.231  & 0.002  \\
 4398.3897 &  -4.6 & 2.8 & -11.3 & 17.4 & 0.235  & 0.001  \\
 4399.3311 &  -1.8 & 2.3 & -28.4 & 22.4 & 0.230  & 0.001  \\
 4591.7121 &   0.6 & 3.3 &  12.9 & 17.4 & 0.234  & 0.001  \\
 4608.7111 &  -4.7 & 3.6 &  20.6 & 21.0 &        &        \\
 4610.7073 &  -9.5 & 3.8 &  39.2 & 19.9 & 0.242  & 0.001  \\
 4630.6769 &  -7.1 & 3.4 & -35.0 & 20.1 & 0.241  & 0.001  \\
 4637.6663 & -13.4 & 3.8 & -14.1 & 19.9 & 0.233  & 0.001  \\
 4664.6389 &   4.2 & 5.3 & -40.4 & 25.7 & 0.255  & 0.002  \\
 4669.6213 & -24.3 & 4.6 & -57.5 & 24.0 & 0.236  & 0.002  \\
 4670.6135 &  -9.6 & 3.6 & -38.9 & 17.5 & 0.228  & 0.001  \\
 4693.6365 &  -6.5 & 3.5 & -49.1 & 20.9 & 0.232  & 0.002  \\
 4694.6189 & -11.5 & 4.7 & -16.0 & 26.3 & 0.233  & 0.002  \\
 4783.4261 &  13.9 & 7.3 &  -9.5 & 30.3 & 0.225  & 0.002  \\
 4819.3328 &   5.4 & 5.0 &  -2.5 & 20.8 & 0.241  & 0.001  \\
 4961.6367 &  -8.0 & 3.2 & -36.9 & 14.2 & 0.238  & 0.001  \\
 4962.6602 &  -6.8 & 2.3 & -36.6 & 15.8 &        &        \\
 5046.4688 &   5.4 & 6.8 & -31.3 & 26.4 &        &        \\
 5096.4835 & -10.5 & 5.2 &  11.4 & 22.4 & 0.244  & 0.002  \\
 5163.4244 &  -6.3 & 4.2 &  50.9 & 15.6 & 0.232  & 0.001  \\
 5193.3521 & -18.3 &10.1 &  -6.3 & 32.6 & 0.232  & 0.002  \\
 5432.4556 &  -4.3 & 7.9 &  16.1 & 29.1 & 0.247  & 0.002  \\
 5434.4336 &   2.8 & 5.2 & -38.1 & 22.4 & 0.207  & 0.001  \\
 5494.3927 &   5.5 & 6.5 &  24.8 & 28.4 & 0.264  & 0.002  \\
 5517.3005 &  -5.3 & 3.9 & -13.8 & 19.3 & 0.240  & 0.001  \\
 5554.3741 & -16.0 & 9.5 &  29.6 & 29.4 & 0.255  & 0.002  \\
 5643.7446 &  -2.6 & 8.6 & -17.0 & 29.3 & 0.271  & 0.002  \\
 5731.5764 & -11.4 & 6.8 &  12.0 & 25.8 &        &        \\
 5741.6407 & -16.9 & 3.0 & -46.4 & 11.0 &        &        \\
 5742.7334 &  -5.9 & 4.7 &  -6.4 & 20.3 & 0.252  & 0.001  \\
 5758.5391 & -12.7 & 5.0 &       &      & 0.258  & 0.001  \\
 5790.5596 & -26.4 & 6.1 &       &      & 0.254  & 0.002  \\
 5801.5615 &  -4.3 & 5.6 &       &      & 0.244  & 0.001  \\
 5807.6191 &   7.6 & 5.0 &       &      & 0.259  & 0.001  \\
\noalign{\smallskip}
\hline
\end{tabular}
\end{table*}

\end{document}